\documentclass[aps,nofootinbib,notitlepage,prd,twocolumn,
groupedaddress,superscriptaddress]{revtex4}

\usepackage{epsfig,amsmath,amsfonts,amssymb,amsthm,graphicx,color,subfigure}

\newcommand{\dm}[1]{\ensuremath{\Delta m^2_{#1}}}
\newcommand{\delCP}{\delta_{\rm CP}}
\newcommand{\sinsq}[1]{\ensuremath{\sin^2\theta_{#1}}}
\newcommand{\chimin}{\chi^2_{\rm min}}
\newcommand{\beq}{\begin{equation}}
\newcommand{\eeq}{\end{equation}}
\newcommand{\beqa}{\begin{eqnarray}}
\newcommand{\eeqa}{\end{eqnarray}}
\newcommand{\barr}{\begin{array}}
\newcommand{\earr}{\end{array}}
\newcommand{\bit}{\begin{itemize}}
\newcommand{\eit}{\end{itemize}}
\newcommand{\nn}{\nonumber}

\def\Esat{E_\mu^{\rm Sat}}
\def\Eopt{E_\mu^{\rm opt}}
\def\gs{\mathrel{
   \rlap{\raise 0.511ex \hbox{$>$}}{\lower 0.511ex \hbox{$\sim$}}}}
\def\ls{\mathrel{
   \rlap{\raise 0.511ex \hbox{$<$}}{\lower 0.511ex \hbox{$\sim$}}}}

\begin{document}

\title{Optimization of the baseline and the parent muon energy \\ 
for a low energy neutrino factory}

\author{Amol Dighe}
\email{amol@theory.tifr.res.in}
\affiliation{Tata Institute of Fundamental Research, 
Homi Bhabha Road, Colaba, Mumbai 400005, India}

\author{Srubabati Goswami}
\email{sruba@prl.res.in}
\affiliation{Physical Research Laboratory, Navrangpura,
Ahmedabad 380009, India}

\author{Shamayita Ray}
\email{sr643@cornell.edu}
\affiliation{Laboratory for Elementary-Particle Physics, 
Cornell University, Ithaca, NY 14853, USA}

\date{\today}

\begin{abstract}
We discuss the optimal setup for a low energy neutrino factory in
order to achieve a $5\sigma$ discovery of a nonzero mixing angle
$\theta_{13}$, a nonzero CP phase $\delCP$, and the mass hierarchy.
We explore parent muon energies in the range 5--16 GeV, and baselines
in the range 500--5000 km. We present the results in terms of the
reach in $\sin^2\theta_{13}$, emphasizing the dependence of the
optimal baseline on the true value of $\delCP$. We show that the
sensitivity of a given setup typically increases with parent muon
energy, reaching saturation for higher energies. The saturation energy
is larger for longer baselines; we present an estimate of this
dependence. In the light of the recent indications of a large
$\theta_{13}$, we also determine how these preferences would change if
indeed a large $\theta_{13}$ is confirmed. In such a case, the
baselines $\sim 2500$ km ($\sim 1500$ km) may be expected to lead to
hierarchy determination ($\delCP$ discovery) with the minimum
exposure.
\end{abstract}

\maketitle

\section{Introduction} 
\label{sec:intro}

While the indications of neutrino oscillations first came from
the solar and atmospheric neutrino data, the observations from
terrestrial experiments have helped provide a firm footing
to our knowledge of neutrino masses and mixing.
The data from all the neutrino oscillation experiments have 
established that there are two independent mass squared differences
$|\dm{31}| \approx 2.35 \times 10^{-3}$ eV$^2$
and $\dm{21} \approx 7.58 \times 10^{-5}$ eV$^2$, 
as well as two large mixing angles $\sin^2 \theta_{23} \approx 0.42$ and 
$\sin^2 \theta_{12} \approx 0.306$ \cite{latest-fit-lisi,latest-fit-schwetz}.
The third mixing angle $\theta_{13}$ is small: at the $3\sigma$ 
level we have an upper bound $\sin^2 \theta_{13} < 0.044$(0.035) 
\cite{latest-fit-lisi}(\cite{latest-fit-schwetz}). 
Recent indications of a nonzero $\theta_{13}$ have been obtained
at the T2K \cite{t2k-th13} and MINOS \cite{minos-th13}, and 
now the global fits that incorporate these new data give a non-zero 
$\theta_{13}$ at $\sim 3\sigma$. 
However, the precise value of the $\theta_{13}$ best fit point as well 
as the significance for $\theta_{13} > 0$ still depend on assumptions 
on the analysis of data from reactor experiments \cite{latest-fit-schwetz}.
(The more recent results from Daya Bay \cite{dayabay-th13} and
RENO \cite{reno-th13} experiments,
which was announced while this paper was under review, claim more 
than $5\sigma$ discovery of a non-zero $\theta_{13}$.
While our analysis has been done assuming that the value of $\theta_{13}$
is still unknown, the implication of such a large $\theta_{13}$
will be discussed towards the end of the paper.)

The immediate goals for neutrino oscillation experiments are the
measurements of (i) the mixing angle $\theta_{13}$, (ii) the CP 
violating phase $\delCP$, and (iii) the sign of $\dm{31}$, 
also known as the mass hierarchy. 
These three quantities, along with the precision measurements of 
the already known ones, are necessary in order to complete our 
knowledge of the neutrino mass spectrum.

Among the three quantities mentioned above, $\theta_{13}$ is the most
important since the determination of the other two depends crucially
on the value of this parameter. 
If $\theta_{13}=0$, the CP phase is an unphysical
quantity, while the determination of mass hierarchy, though possible
in principle \cite{degouvea-winter}, becomes extremely challenging. 
If $\sin^2 2\theta_{13} \gtrsim 0.01$, its measurement will be within
the reach of accelerator experiments like T2K, MINOS or NO$\nu$A 
that use conventional hadron beams, for certain $\delCP$ values. 
Indeed the recent results \cite{t2k-th13,minos-th13} indicate that such a
measurement may soon be possible.
If on the other hand 
the $\nu_\mu \to \nu_e$ signals observed here are background
or statistical fluctuations, then $\theta_{13}$ may be even smaller.
Reactor experiments, Double CHOOZ, Reno and Daya-Bay may probe this
angle as long as $\sin^2 2\theta_{13} \gtrsim 0.033,0.018, \mathrm{and}~ 0.007$
respectively \cite{reactor-experiments}, 
independent of the value of $\delCP$ .
(The Daya Bay \cite{dayabay-th13} and RENO \cite{reno-th13}
experiments already claim the 
measurement of this angle to more than $5\sigma$, as mentioned earlier.)

The most efficient way of determining mass hierarchy is the observation 
of the difference in Earth matter effects for the two hierarchies,
which is possible if $\theta_{13}$ is sufficiently large.
Future atmospheric neutrino experiments can achieve this task for 
$\sin^2 2\theta_{13} \sim 0.04$ at 95\% C.L. \cite{atm-hierarchy}. 
For T2K and NO$\nu$A, the sensitivity to hierarchy
is possible for $\sin^2 2\theta_{13} > 0.02$ at 90\% C.L.,
albeit only for limited range of values of $\delCP$ 
\cite{lindner-superbeam}.

The measurement of $\delCP$ is perhaps the most difficult of the 
three. Not only does it need a substantial value of $\theta_{13}$,
the value of $\delCP$ itself needs to be sufficiently different
from zero for a positive signal of CP violation. 
Planned superbeam experiments, which would use highly intense
conventional beams, have a limited sensitivity to $\delCP$
at 90\% C.L. and almost no sensitivity at 3$\sigma$ \cite{lindner-superbeam}.

While the technology for the conventional beams is well-established,
the beam contamination inherent in such beams does not allow
measurements accurate to more than a percent level. On the other
hand, neutrino production from the decays of muons that are
accelerated and stored in a ring (``neutrino factory''),
combined with detectors that
can identify the charge of leptons produced from the neutrino
interactions, has the potential of measuring the quantities of
interest even for much smaller $\theta_{13}$ values.
Even if the conventional beams succeed in an measurement, it is
important to confirm such a measurement with another type of source,
just like the measurements from solar and atmospheric experiments
were later confirmed and established by terrestrial neutrino
experiments.
A neutrino factory is thus a discovery machine in the worst-case
scenario (measurements beyond the reach of conventional beams)
and a precision machine in favorable scenarios.

In a neutrino factory, accelerated muons are allowed to decay 
in the long straight sections of a storage ring, resulting in a 
strong collimated beam. 
A $\mu^+$ beam decays to give $\bar{\nu}_\mu$ and $\nu_e$. 
Oscillations of the $\nu_e$ to $\nu_\mu$ produce a $\mu^{-}$ in 
the detector giving the so called ``wrong sign'' muon signal, 
whereas the unoscillated $\bar{\nu}_\mu$ produce the ``right sign''
or ``same sign'' muon signal of $\mu^+$. 
A detector with a charge identification capability can identify the 
two different type of signals separately, allowing the determination
of $P_{e\mu} \equiv P(\nu_e \to \nu_\mu)$ and
$\bar{P}_{\mu \mu} \equiv P(\bar\nu_\mu \to \bar\nu_ \mu)$.
A $\mu^-$ beam would similarly lead to the measurements of
$\bar{P}_{e \to \mu} \equiv P(\bar\nu_e \to \bar\nu_\mu)$ and 
$P_{\mu\mu} \equiv P(\nu_\mu \to \nu_\mu)$. 
While the sensitivity stems mainly from the wrong sign muon 
signal due to the appearance channels $\nu_e \to \nu_\mu$
and $\bar\nu_e \to \bar\nu_\mu$, the disappearance channels
$\nu_\mu \to \nu_\mu$ and $\bar\nu_\mu \to \bar\nu_\mu$ also
contribute because of the large statistics available in these
channels.
The sensitivity of a neutrino factory may be up to $\sin^2 2 \theta_{13}$ 
as low as $\sim 10^{-4}$ \cite{nufact}.

Since the appearance probability $P_{e\mu}$ depends on all the three 
hitherto unknown quantities, the channel $\nu_e \to \nu_\mu$ in principle
contains information on all of them, and hence it has been acclaimed as 
the golden channel.
However this advantage is masked by the fact that the value of $\delCP$ is 
completely unknown. This gives rise to degenerate solutions making the 
unambiguous determination of the oscillation parameters an uphill task. 
An elegant solution was provided by observing that at the distance 
$\sim$7500 km, the golden channel probability $P_{e \mu}$ 
becomes independent of the CP violating phase irrespective of energy, 
hierarchy and oscillation parameters \cite{MAGIC}. 
This baseline is the so called ``magic'' baseline. 
A neutrino factory is considered most suitable for a magic baseline 
experiment because of the large flux and the small beam background.

Of course since the oscillation probability at the magic baseline is 
independent of the CP phase, there is no sensitivity to $\delCP$.
Therefore though the magic baseline experiment is suitable 
for determination of hierarchy and $\theta_{13}$, one has to consider
other shorter baselines for the $\delCP$ determination.
Detailed energy-baseline optimization studies by the IDS-NF group propose
two magnetized iron neutrino detectors (MIND), one at a distance of 4000 km 
and another at the distance of 7500 km, 
with a muon energy of 25 GeV \cite{ids-nf}. 
However this requires high acceleration of the muons, and 
one has to contend with the $1/r^2$ fall-off of the flux.

In the recent past many authors have investigated the prospect of having 
a neutrino factory of much lower energy (4 -10 GeV), and hence a shorter 
baseline. This was termed as the low energy neutrino factory (LENF) 
\cite{lenf,lenf-improve}. 
The preferred detectors at these low energies are the magnetized totally 
active scintillator detectors (TASD) or Liquid Argon detectors that 
can detect muon change efficiently.
Recently the MIND-type detectors for LENF were also considered
\cite{agarwalla-optimal}.
Non-magnetic detectors for LENF has been considered in \cite{huber-lenf}.

The most discussed baseline in the context of LENF is the
DUSEL baseline of 1300 km. 
However recently it was pointed out in \cite{2540-umasankar} that the baseline 
$\sim$ 2540 km has a special property that in the inverted hierarchy (IH),
the golden channel probability $P_{e\mu}$ is independent of the CP phase 
around 3.3 GeV and hence it can be used for an efficient determination 
of hierarchy using the $\nu_\mu \to \nu_e$ channel in superbeams. 
In a subsequent Letter \cite{dgrprl}, we showed that for the same baseline, 
the probability $P_{e\mu}$ in the normal hierarchy (NH) also becomes 
independent of the CP phase at 1.9 GeV. 
Therefore we termed this as the ``bimagic'' baseline and the energies as 
the magic energies. 
We also observed that away from the magic energies, the probabilities 
still depend on $\delCP$. Therefore an experiment at this baseline 
would be sensitive to all the three parameters if one uses a broadband 
neutrino beam from 1--4 GeV as can be obtained from a 5 GeV neutrino 
factory. 
A noteworthy point is that the distance 2540 km which was motivated purely from 
physics considerations in \cite{2540-umasankar} and \cite{dgrprl} 
happens to be close to the Brookhaven--Homestake \cite{bnlhomestake} 
and CERN--Pyh\"{a}salmi \cite{lena} baselines \footnote{ 
Potential of baselines close to 2540 km has been studied in 
\cite{bimagic-superbeam} in the context of superbeams.}. 

We note that the LENF was initially motivated to study precision 
neutrino properties at large $\theta_{13}$. Therefore, 
with the current indication that $\theta_{13}$ is indeed large, 
an exploration of the potential of a LENF is worthwhile and timely.
Since the neutrino-factory technology is still not well-established,
the time scale at which these experiments will start is comparatively
larger, and therefore its aim should be correspondingly higher -- like
the measurements of the above quantities of interest to $5\sigma$.
This is particularly true for a quantity like the mass hierarchy,
which is a binary measurement.
A large $\theta_{13}$ is also conducive for a measurement of $\delCP$ 
and it is likely that this aspect may play a decisive role in ascertaining 
which is the optimal baseline and energy. 
Of course if $\theta_{13}$ happens to be smaller then it needs to be explored 
what is the optimum baseline and energy for determination of the all the three 
unknowns -- mass hierarchy, $\theta_{13}$ and $\delCP$. 

In general, optimization is a complex numerical problem. The parameters 
involved are energy, baseline as well as the true values of the oscillation 
parameters. There is also the issue of the optimization with
respect to the detector. 
The dependence of probabilities on the true values of $\theta_{13}$, 
$\delCP$ and $\dm{31}$ is beyond the experimental control, so
the reach of an experiment has to be assessed in the worst case
scenario as far as the values of the mixing parameters are concerned.
As we shall show further in this paper 
the lack of knowledge of true $\delCP$ makes it extremely difficult 
to zero in on a particular baseline as ``the optimal baseline'' using 
the so called green-field approach, where one determines the optimal 
baseline by a numerical scan of the relevant parameter space. 

Optimization studies in the context of low energy neutrino factories 
have been carried out recently in \cite{agarwalla-optimal} 
in the context of a MIND.
In this paper we consider a detector that is TASD.
Apart from this, one of the major differences in their analysis and 
ours is that they present the sensitivity plots 
in terms of ``fraction of $\delCP$'', whereas our 
results are presented for all $\delCP$ values in the range
$[0 - 2 \pi]$ and our plots reveal the specific range of values 
of $\delCP$ for which a given baseline is 
sensitive to a particular quantity.

The plan of the paper goes as follows. In Sec.~\ref{probs} we discuss the 
physics of oscillation probabilities giving us the bimagic condition and 
the deviations from this condition in the nearby baselines. 
In Sec.~\ref{setup} we discuss the experimental set up and the details of 
the numerical simulation. 
We present the results for the baseline optimization in Sec.~\ref{sec:Lopt} 
and those for the muon energy optimization in Sec.~\ref{sec:Eopt}.
The dependence on the true values of $\delCP$ and $|\dm{31}|$ is shown
by bands obtained by varying these parameters over their currently allowed 
ranges. The estimation of an optimal muon energy, given a baseline, is 
outlined in Sec.~\ref{emu-given-baseline} with a simple approximation.
In Sec.~\ref{large-t13} we discuss the implications of a large measured
$\theta_{13}$ value for the optimization.
Sec.~\ref{concl}, we summarize our results and comment on future prospects.

\section{Magic and bimagic baselines}
\label{probs}

Certain properties of the neutrino flavor conversion probability
$P_{e\mu}$ can be useful for an analytical understanding of why
certain baselines or parent muon energies should work better than
the others. While these have been pointed out earlier 
\cite{smirnovmagic,2540-umasankar,dgrprl}, we expound on them here
in detail, bringing out some of their most important features.

In general the $P_{e \mu}$ oscillation probability can be written as
\cite{smirnovmagic} 
\beq
P_{e \mu} = |\cos\theta_{23} A_S e^{i \delCP}  + \sin\theta_{23} A_A|^2 
\eeq 
where $A_S$ is the ``solar'' amplitude that depends on the solar parameters 
$\dm{21}$ and $\theta_{12}$, and $A_A$ is the ``atmospheric'' amplitude which 
depends on $\dm{31}$ and $\theta_{13}$. From the above expression it 
is evident that the CP violation in neutrino oscillation arises from the 
interference effects of these two amplitudes. 
In matter of constant density, 
the oscillation probability $P_{\nu_e \to \nu_\mu}$ can be expanded
keeping terms upto second order in the small parameters 
$\alpha \equiv \dm{21}/\dm{31}$ and $s_{13}$
as  
\cite{akhmedov-prob}
\beqa
P_{e \mu}&=&4 s_{13}^2 s_{23}^2 \dfrac{\sin^2{[(1 -\hat A)\Delta]}}{(1-\hat A)^2} 
+ \alpha^2 \sin^2{2\theta_{12}} c_{23}^2 \dfrac{\sin^2{\hat A \Delta}}{\hat A^2} \nn \\ 
&& +  2 \alpha s_{13}  \sin{2\theta_{12}} \sin{2\theta_{23}} \cos{(\Delta - \delCP)} \times \nn \\
& & \phantom{space} \dfrac{\sin{\hat A \Delta}}{\hat A} \dfrac{\sin{[(1-\hat A)\Delta]}}{(1-\hat A)} \; ,
\label{P-emu}  
\eeqa
where $s_{ij} \equiv \sin{\theta_{ij}}$, $c_{ij} \equiv \cos{\theta_{ij}}$. Also,
\beqa
\hat A \equiv \dfrac{2\sqrt{2} G_F n_e E_\nu}{ \dm{31}} \; , \quad 
\Delta \equiv \dfrac{\dm{31} L }{4 E_\nu} \; ,
\label{Ahat-Delta}
\eeqa
where $G_F$ is the Fermi constant and $n_e$ is the electron number density.
For neutrinos, the signs of $\hat{A}$ and $\Delta$ are positive for 
normal hierarchy and negative for inverted hierarchy. 
$\hat{A}$ picks up an extra negative sign for anti-neutrinos.
The last term in Eq.~(\ref{P-emu}) corresponds to the interference term 
from which the CP dependence of the probability originates. 
This term also mixes the dependence on hierarchy and $\delCP$, as well as 
the dependence on $\theta_{13}$ and $\delCP$, leading to a four-fold 
degeneracy \cite{intrinsic}.
There is also a degeneracy between ($\theta_{23}, \delCP$) and
($\pi/2 - \theta_{23},\delCP$) \cite{lisi-degen}.
Together, this eightfold degeneracy 
makes the determination of the oscillation parameters ambiguous. 
It was noticed in \cite{MAGIC} that the CP dependence of the probability 
can be avoided if one has 
\beq
\sin(\hat A \Delta) = 0 \; ,
\label{solar-amp} 
\eeq 
which corresponds to the vanishing of the solar amplitude $A_S$ and hence 
the interference term. As a result $P_{e \mu}$ becomes 
independent of the CP phase $\delCP$ as well as the solar parameters. 
This condition is obeyed at the so-called magic baseline 
($L \sim 7500$ km) for all $E_\nu$ and for both the hierarchies.

\begin{figure*}[]
\begin{center}
\includegraphics[width=0.9\columnwidth]{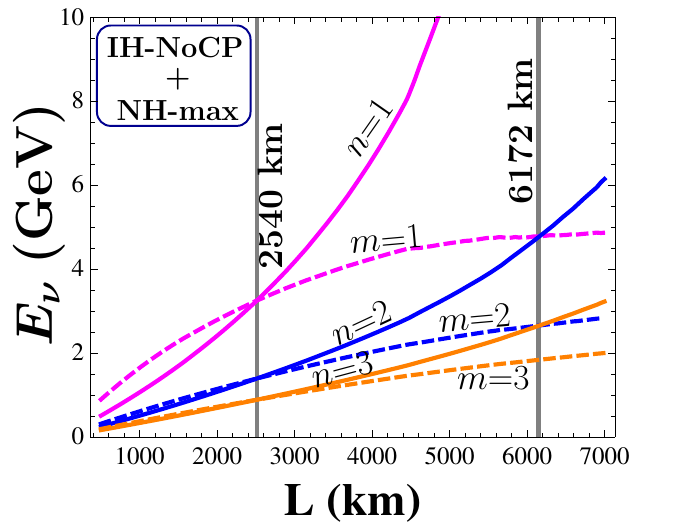}
\includegraphics[width=0.9\columnwidth]{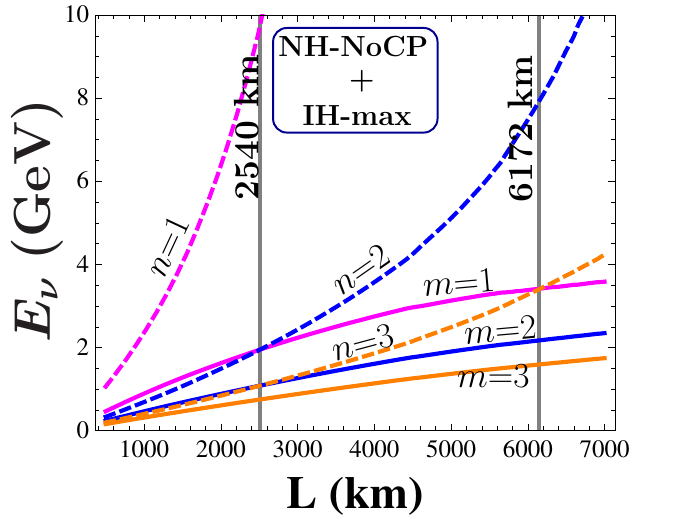}
\caption{(Color online) Graphically solving the pair of magic conditions.
Left panel: $E_{\rm magic}^{\rm IH}$ (solid) and $E_{\rm max}^{\rm NH}$ (dashed)
as a function of $L$ for different $n,m$.
Right panel: $E_{\rm magic}^{\rm NH}$ (solid) and $E_{\rm max}^{\rm IH}$ (dashed)
as a function of $L$ for different $n,m$.
The intersection points of the curves show the baselines and the
energies at which the conditions for maximum hierarchy sensitivity are
satisfied.}
\label{bimagic}
\end{center}
\end{figure*}


However, the dependence on the CP phase also vanishes when the atmospheric 
amplitude $A_{A}$ vanishes \cite{smirnovmagic}, which corresponds 
to the condition 
\beq 
\sin[(1-\hat A)\Delta] = 0 \; . 
\label{atm-amp} 
\eeq
However one notes that unlike the magic baseline condition this
condition depends on energy as well as hierarchy. 
Therefore for a particular hierarchy and a particular baseline one can find 
a set of magic energies where the CP dependence in the probability vanishes. 
If we consider inverted hierarchy (IH), condition for no $\delCP$ 
sensitivity (IH-NoCP) can be written as 
\beq
(1+ |\hat A|) \cdot |\Delta| = n \pi \; 
\label{atm-amp-ih}
\eeq
with the integer $n > 0$. The magic energies are given as 
\beq
E_{\rm magic}^{\rm IH} = \dfrac{1.27 |\dm{31}| L}{n\pi - 1.27 \;|K| \rho L } \; .
\label{emagicih} 
\eeq 
Here the quantity $K$ has been defined such that 
$K \rho \equiv 2\sqrt{2} G_F n_e$, where $\rho \equiv \rho(L)$ 
denotes the average matter density for the baseline $L$.
For a given baseline $L$, at these energies the probability $P_{e \mu}$ for IH 
is independent of the CP phase as well as $\theta_{13}$ and 
only the ${\cal{O}}(\alpha^2)$ term contributes. 
The hierarchy dependence of the magic energy can be utilized to maximize the 
hierarchy sensitivity by demanding that $\sin[(1-\hat A)\Delta] = \pm 1$ 
for normal hierarchy (NH) at the same time. 
We will refer this condition as NH-max.
With this condition, the number of events are enhanced due to the 
first term in Eq.~(\ref{P-emu}). Also the CP dependence is retained 
in the NH probability. The condition for maxima in NH can be expressed as 
\beqa 
(1- |\hat A|)\cdot |\Delta| & = &  (m -1/2) \pi \; ,
\label{nh-1}
\eeqa
when $m$ is any integer. This gives
\beqa
E_{max}^{\rm NH} = 
\dfrac{1.27 |\dm{31}| L}{(m-1/2)\pi + 1.27 \; |K| \rho L} \; .
\label{emaxnh}
\eeqa
It may be conjectured that we will have maximum hierarchy sensitivity if 
\beq
E_{\rm magic}^{\rm IH} = E_{\rm max}^{\rm NH} \; .
\label{eq:bimagic}
\eeq
From the condition in Eq.~(\ref{eq:bimagic}) we arrive at the baseline 
\begin{equation}
\rho L {\rm (km \ g/cc)}  \approx  (n-m + 1/2) \times  16300 \; . 
\label{rhol-ihnocp} 
\end{equation}
Note that the relevant $L$ in Eq.~(\ref{rhol-ihnocp}) is independent of 
any oscillation parameters as in the case of the magic baseline. 
However unlike the magic baseline this condition will be satisfied only
for particular values of energy, given by 
$E_{\rm magic}^{\rm IH} = E_{\rm max}^{\rm NH}$.

Alternatively if we demand no sensitivity to CP phase in NH (NH-NoCP), we get
\beqa
(1 - |\hat A|)\cdot |\Delta| & = &  n \pi  \; ,  
\label{nh-0}
\eeqa
which gives 
\beq
E_{\rm magic}^{\rm NH} = \dfrac{1.27 |\dm{31}| L}{n\pi + 1.27 \;|K| \rho L} \; ,
\label{emagicnh}
\eeq 
where $n$ is any non-zero integer. 
For a given baseline $L$, at these energies the probability 
$P_{e \mu}$ for NH is independent of the $\delCP$ as well as $\theta_{13}$, 
and $P_{e \mu}$ becomes ${\cal{O}}(\alpha^2)$.
The condition for maxima for IH (IH-max) gives
\beqa 
(1 + |\hat A|) \cdot |\Delta| & = &  (m - 1/2) \pi \; ,
\label{ih-1}
\eeqa
where $m$ is a positive integer. This gives 
\beqa
E_{\rm max}^{\rm IH} = 
\dfrac{1.27 |\dm{31}| L}{(m-1/2)\pi - 1.27 \; |K| \rho L } \; .
\label{emaxih}
\eeqa
Demanding 
\beq
E_{\rm magic}^{\rm NH} = E_{\rm max}^{\rm IH} 
\label{eq:bimagic-2}
\eeq 
we get the same equation for the baseline $L$, as in Eq.~(\ref{rhol-ihnocp}), 
except for an overall negative sign which can be attributed to the 
different regions of validity for $n,m$.  

\begin{figure*}[]
\begin{center}
\includegraphics[scale=0.55]{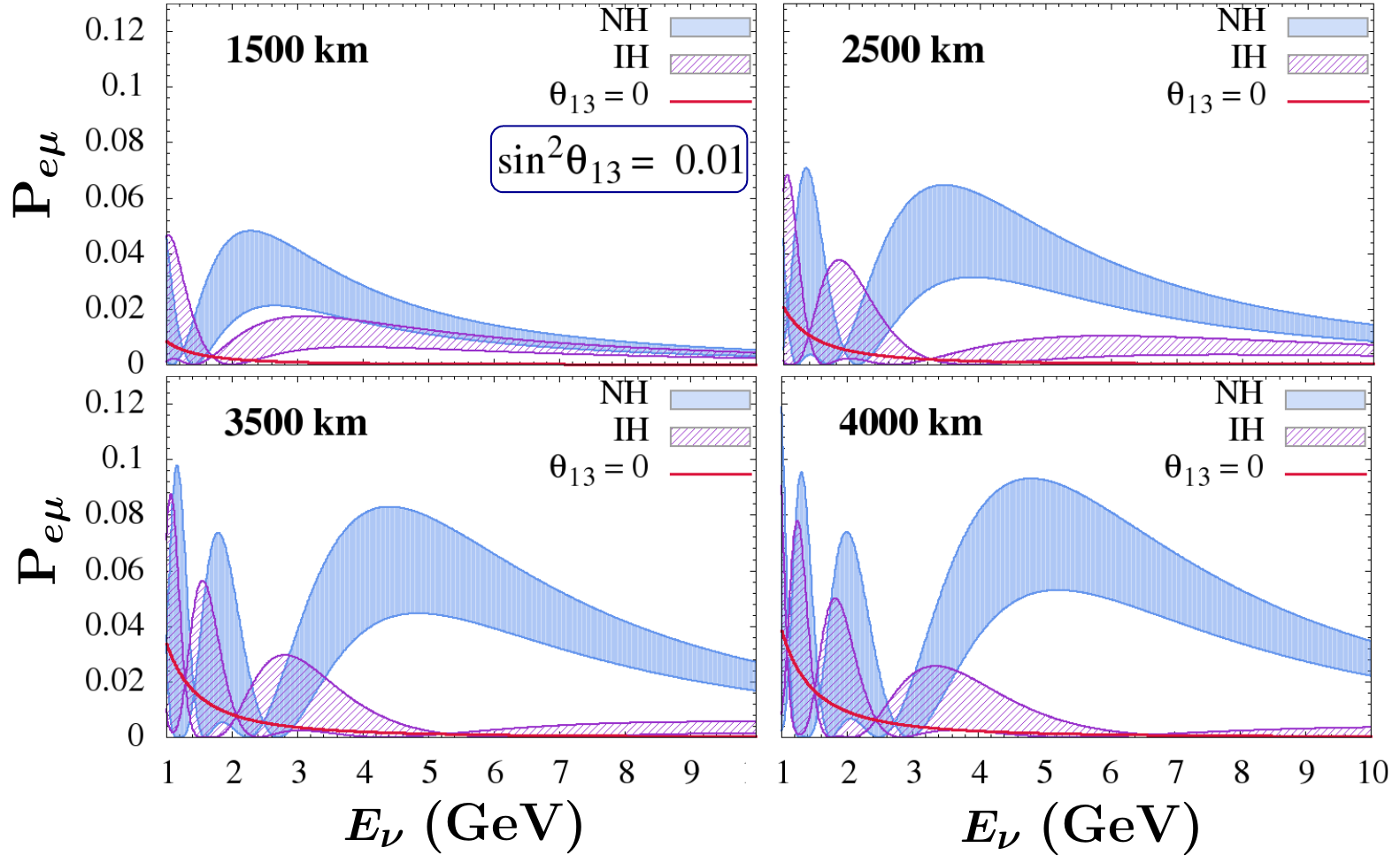}
\caption{(Color online) Conversion probability $P_{e \mu}$ for $L$ = 1500 km, 
2500 km, 3500 km and 4000 km. The bands correspond to 
$\delCP \in [0, 2\pi]$. Other parameters are as given in Eq.~(\ref{simulation-para}).
The red (thick) line corresponds to $\theta_{13}$ = 0.
\label{fig:prob}}
\end{center}
\end{figure*}


Fig.~\ref{bimagic} demonstrates the existence of bimagic baselines 
along with the corresponding magic energies, for $n,m = 1,2,3$.
The left panel shows the solutions for IH-NoCP and NH-max, while 
the right panel shows NH-NoCP and IH-max solutions.
It is clear from the figure that the baseline $L \sim 2540$ km 
is the shortest baseline that satisfies both the pairs of conditions 
simultaneously. It is not a surprise, since the equation for the baseline 
Eq.~(\ref{rhol-ihnocp}) is similar for both the pairs of conditions.
This baseline is therefore termed as bimagic \cite{dgrprl}.
For the pair IH-NoCP and NH-max, the solution is obtained 
from Eq.~(\ref{rhol-ihnocp}) with $n=m$ and an average matter 
density of $\rho \sim 3.2$ g/cc. 
The corresponding magic energies can be obtained from Eq.~(\ref{emagicih}) 
or Eq.~(\ref{emaxnh}) to be 3.3 GeV for $(n,m)=(1,1)$, 1.4 GeV for 
$(n,m)=(2,2)$, 0.89 GeV for $(n,m)=(3,3)$ and so on. 
Clearly the magic energies decrease with increasing $n,m$.
Larger values of $n,m$ are less and less practical since the flux
at low energies, as well as the efficiency of detection, is typically lower.
The solution for the pair NH-NoCP and IH-max is similarly obtained,
the magic energies here are 1.97 GeV ($n=1,m=2$) and 1.09 GeV ($n=2,m=3$).

As can be seen from the figure, the baseline of 6172 km also satisfies 
the two pairs of conditions.
With $\rho(6172~km) = 3.955$ g/cc, IH-NoCP and NH-max are satisfied for
$n-m=1$; the magic energies in the relevant range are 4.8 GeV for 
$(n,m)=(2,1)$ and 2.66 GeV for $(n,m)=(3,2)$.
For NH-NoCP and IH-max, the interesting energy is 3.42 GeV,
obtained with $(n,m)=(1,3)$.
This baseline therefore also deserves the title ``bimagic''.
So do the longer baselines of 8950 km and 10690 km, which are not
shown in the figure. 
However for the purpose of numerical optimization studies in this paper,
we restrict ourselves to baselines in the range 500-5000 km,
since longer baselines imply a lower flux, 
following the $1/r^2$ behavior.

Even though solving the bimagic conditions given in 
Eqs.~(\ref{eq:bimagic}) and (\ref{eq:bimagic-2}) using the 
Preliminary Reference Earth Model (PREM) \cite{prem} 
profile we get the exact values 
of the bimagic baselines with the magic energies, 
it is observed from Fig.~\ref{bimagic} that 
the values of magic energy for one hierarchy and the 
maximum energy for the other hierarchy move away from 
each other rather slowly on either sides of the intersection point, 
as $L$ is varied. 
Moreover, currently there is $\sim$5\% error on $|\dm{31}|$ at 1$\sigma$ 
as well as uncertainties associated with the density profile of Earth.
So the excellent hierarchy sensitivity of 
the bimagic baselines, attributed to its bimagic property,
is expected to be there even if we move slightly away 
from these specific baselines. 
There already exist a few 
possible baselines of similar magnitudes:
(i) Brookhaven to Homestake distance is exactly 2540 km, and
(ii) CERN to Pyh\"{a}salmi (proposed site for the LENA detector) 
distance is 2288 km.

Fig.~\ref{fig:prob} shows the probability $P_{e \mu}$ for different
baselines, for $\sinsq{13}=0.01$.
In this and all other plots,
we have solved the exact neutrino propagation equation numerically 
using the PREM profile. 
The probability $P_{e\mu}$ shown in Fig.~\ref{fig:prob} shows the presence 
of the bimagic properties for $L \sim 2500$ km clearly. 
Around this distance, 
the maximum of NH appears at the same energy as the minimum 
of IH and vice versa, enhancing the hierarchy sensitivity. 
This particular feature is absent in any of the other 
baselines shown. 
When one goes to a higher baseline, the following effects occur: \\
(i) The flux decreases as $1/r^2$.\\
(ii) The amplitude of oscillations in NH increases while that 
in IH decreases. This would tend to increase the hierarchy sensitivity. \\ 
(iii) The oscillation maxima of IH move faster to higher energies than 
that of NH, resulting in the maxima of both hierarchies coming closer 
in energies, which would tend to decrease the hierarchy sensitivity. \\
(iv) At a given parent muon energy, the number of events depends
on the overlap of the flux spectrum and the probability $P_{e\mu}$.
While the flux spectrum is independent of the baseline, the major
maxima in $P_{e\mu}$ (the one at the highest energies) shifts to higher
energies. This results in a decrease in the overlap between the
flux spectrum and $P_{e\mu}$, leading to a decrease in the statistics. \\ 
The net result is a subtle combination of all these effects. 
At the lower end of the parent muon energies $E_\mu$, the bimagic baseline 
gives the best sensitivity to hierarchy, while at higher $E_\mu$ the 
optimal baseline increases. 

The figure also shows the probability for $\theta_{13}$ = 0. 
This is the same for both the hierarchies since in 
Eq.~(\ref{P-emu}) at $\theta_{13}$=0 only the ${\cal O}(\alpha^2)$ 
term contributes in the leading order which is independent of hierarchy. 
The distance of the bands from this line gives an direct estimate 
of $\theta_{13}$ discovery potential for the corresponding baseline 
and hierarchy. The 2500 km plot in Fig.~\ref{fig:prob} suggests that 
$\theta_{13}$ discovery potential is expected to be good for NH, 
while it may not be so good for IH as the $P_{e\mu}$ values are lower.
The same is true 
for the longer baselines. The conclusions get reversed for 
anti-neutrinos, because of the change in sign in $\hat A$ in Eq.~(\ref{P-emu}).
However for 1500 km, the difference between the probability values for 
NH and IH is not so large, and hence $\theta_{13}$ discovery potential 
is expected to be similar for both the hierarchies.
Of course on top of the probabilities, the cross section and the
$1/r^2$ flux dependence also contribute in determining the 
optimal baseline.

The sensitivity to the CP phase is related to the widths of the 
bands that represent the variation of the CP phase. 
The widths of the $P_{e\mu}$ bands seem to increase with increasing baselines.
On the other hand, the fluxes fall as $1/r^2$. So the optimal
baseline should emerge from a compromise between these two opposing
factors.

In this section we have given analytic arguments to motivate the
desirable values for the baselines and neutrino energies
to determine the sign of mass hierarchy, and detect nonzero 
values for $\theta_{13}$ and $\delCP$.
To choose the optimal baseline for the low energy neutrino factory, 
one will have to perform a complete numerical study, which we do in the next 
sections.

\section{Details of numerical simulations} 
\label{setup}

As the detector, we use a 25 kt TASD with an energy threshold of 1 GeV.
We choose a typical Neutrino factory setup with 
$5 \times 10^{21}$ useful muon decays per year, which is of the same
order as in the setup considered in \cite{huber, lenf-improve}.
We consider the running with both the polarities, each for 2.5 years. 
So we have a neutrino flux consisting of $\bar \nu_\mu$ and $\nu_e$ when 
we consider running with $\mu^+$, while it becomes $\nu_\mu$ and $\bar \nu_e$ 
when running with negative polarity.

We assume a muon detection efficiency of 94\% for energies above 1 GeV,
10\% energy resolution for the whole energy range and
a background level of $10^{-3}$ for the $\nu_e \to \nu_\mu$ and
$\bar \nu_\mu \to \bar \nu_\mu$ channels.
Detection of $\nu_e$ or $\bar \nu_e$ is not considered in this study,
which seems to have a very small effect when the initial flux is as 
large as above \cite{lenf-improve}. A 2.5\% normalization error and 
0.01\% calibration error, both for signal and background, have also 
been taken into account throughout this study.
The detector characteristics have been simulated with GLoBES \cite{globes}.
We also use the prescriptions for priors and marginalization in-built in
GLoBES.

Our main goal is to find out the optimal baseline 
as well as the optimal parent muon energy $E_\mu$ for that baseline.
In order to optimize the baseline, we choose three representative 
energies for the parent muon: 5 GeV, 7.5 GeV and 10 GeV,
and vary the baseline in the range 500 -- 5000 km, 
For optimizing the parent muon energy $E_\mu$, we choose 
three representative baselines: 1500 km, 2500 km and 3500 km,
and vary $E_\mu$ in the range 2 -- 16 GeV.

\section{Optimization of the baseline} 
\label{sec:Lopt}

In this section, we present the results of our baseline optimization 
for the measurement of the three ``performance indicators'' \cite{agarwalla-optimal}:
neutrino mass hierarchy, discovery of $\theta_{13}$ and $\delCP$. 
We have performed the analysis at the 
parent muon energies of $E_\mu$ = 5, 7.5 and 10 GeV.
The results at intermediate energies can be extrapolated from the
results at these representative parent muon energies. 
The energy optimization will be presented in the next section.

The main sources of uncertainty in determining the reach of
an experiment are the unknown values of $\delCP$ and $\theta_{13}$. 
We therefore focus on the influence 
of these two quantities on our results, and keep the true 
values of other mixing parameters to be fixed at
\beqa
\Delta m^2_{21} &=& 7.65 \times 10^{-5} \, {\rm eV}^2 \; , \quad 
\sin^2\theta_{12} = 0.3 \; , \nn \\
|\Delta m^2_{31}| &=& 2.4 \times 10^{-3} \, {\rm eV}^2 \; , \quad 
\sin^2\theta_{23} = 0.5 \;.
\label{simulation-para}
\eeqa
We also explore the effects of varying $|\Delta m^2_{31}|$ in its current
$3\sigma$ allowed range.
In this section we present three kinds of plots:  
\\
(i) Type-A: plots in the $\sin^2\theta_{13}-L$ plane for fixed true values of
{$|\Delta m^2_{31}|$} and $\delCP$ varying in the range [0, $2\pi$], \\
(ii) Type-B: plots in the $\delCP- L $ plane for fixed true values of 
$\sin^2\theta_{13}$ and {$|\Delta m^2_{31}|$}, \\
(iii) Type-C: plots in the $\sin^2\theta_{13}-L$ plane for fixed true values of 
$\delCP$ and varying {$|\Delta m^2_{31}|$} in its current 3$\sigma$ range.

In the conventional plots 
the reach for a particular performance indicator is often 
given in terms of the fraction of $\delCP$ values for which the
determination of a quantity is possible.
Our Type-A plots show the reach for all possible $\delCP$ 
values, from which the reach even for the worst-case $\delCP$ values
may be inferred. Moreover, from the Type-B plots that show the reach
for all $\delCP$ values, one may also trivially infer the fraction of 
$\delCP$ for which the quantity may be determined. 

For reference, in the baseline optimization plots we also show two
vertical lines, corresponding to the baselines of 2540 km, the
``bimagic'' baseline, and 1300 km, the baseline that is perhaps
the most studied in the context of the LENF.

\subsection{Hierarchy determination} 
\label{sec:hierarchy-Lopt}

In order to optimize the baseline for the determination of hierarchy,
we assume NH to be the true hierarchy and show the reach of $\theta_{13}$ 
for which the wrong hierarchy (IH) can be excluded to $5\sigma$,
as a function of the baseline.
Note that the determination of hierarchy is a binary measurement, 
and hence a $5\sigma$ determination is absolutely necessary before claiming 
a positive identification of this quantity.

\begin{figure*}[]
\begin{center}
\includegraphics[angle=0,width=1.8\columnwidth]{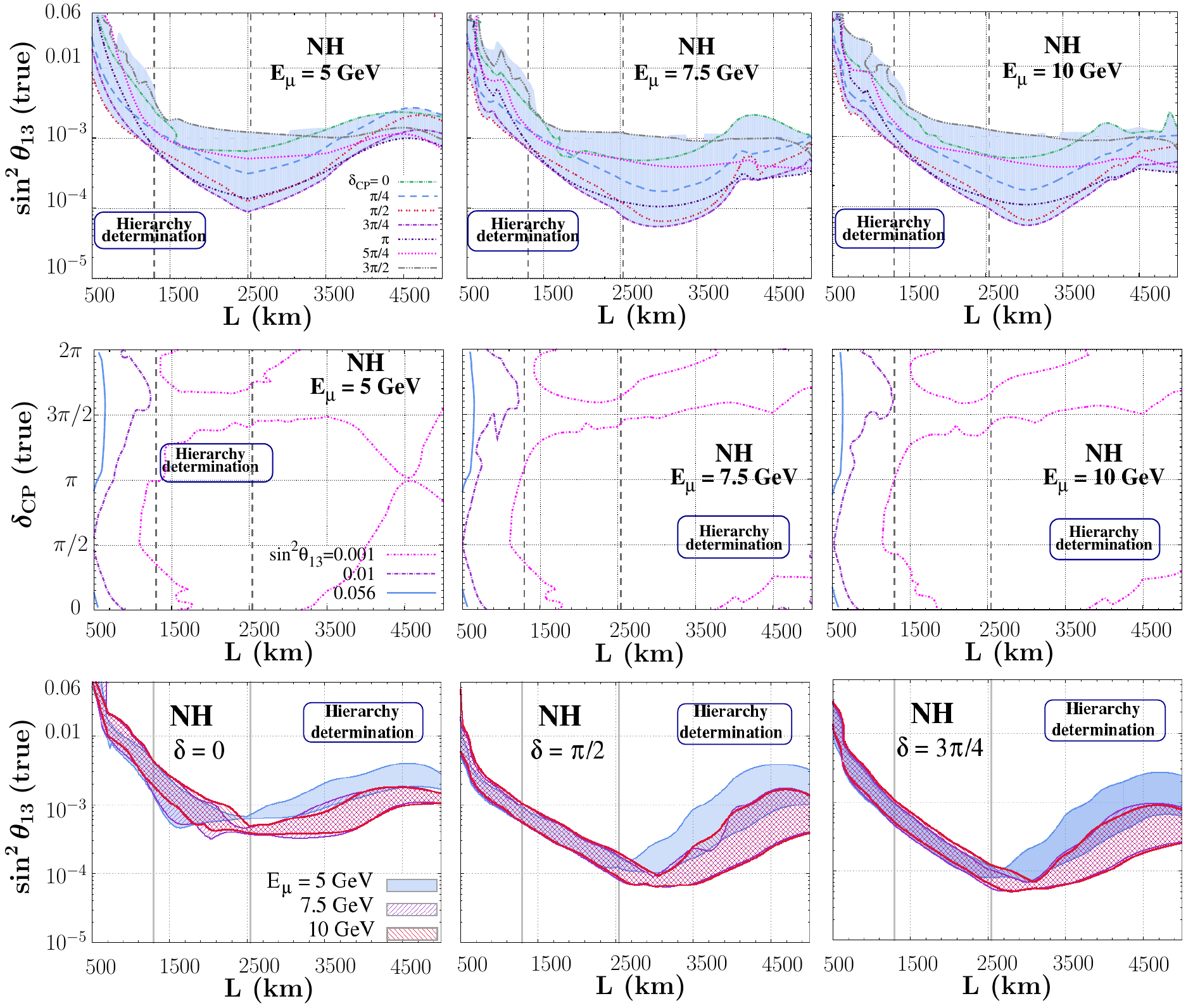} 
\caption{(Color online) 5$\sigma$ reach in hierarchy determination 
for fixed muon energies, as a function of baseline, assuming the true 
hierarchy to be NH. The top panel gives the reach in $\sin^2\theta_{13}$,
the bands correspond to $\delCP \in [0, 2\pi]$. 
Specific values of $\delCP$ are also shown within the band. 
The middle panel shows the 5$\sigma$ reach in $\delCP$ 
for fixed values of $\sin^2\theta_{13}$. The bottom panel denotes the 
reach in $\sin^2\theta_{13}$ for fixed values of $\delCP$. 
The bands correspond to the current 3$\sigma$ range of $|\dm{31}|$.
All the undisplayed parameters are fixed at values given 
in Eq.~(\ref{simulation-para}). 
The plots are generated for 2.5 years of running with each muon polarity.
The two dark vertical lines correspond to the baselines of 2540 km (the
bimagic baseline), and 1300 km.
}
\label{hierarchy-Lopt}
\end{center}
\end{figure*}

Figure~\ref{hierarchy-Lopt} shows the hierarchy sensitivity 
with the stated experimental setup. 
In the top panel we present Type-A plots, where 
true values of $\sinsq{13}$ are plotted along the vertical 
axis. True values of all other parameters, except $\delCP$, 
are set to values stated in Eq.~(\ref{simulation-para}).
To generate the bands in this top panel,
$\delCP$(true) is varied over the complete range of $[ 0, 2\pi]$. 
For each set of chosen values of mixing parameters and chosen baseline, 
$\chimin$ is obtained by marginalizing over all parameters with 
the wrong hierarchy. We have taken 4\% error on each of $\dm{21}$ and 
$\theta_{12}$ and 5\% on $\theta_{23}$ and $|\dm{31}|$. 
A 2\% error has also been
considered on the Earth matter profile and marginalized over.

The figures in the top panel of 
Fig.~\ref{hierarchy-Lopt} show that the baseline optimization 
depends on the actual $\delCP$ value as well as the
parent muon energy $E_\mu$. The widths of the bands, which span almost
an order of magnitude in $\sin^2\theta_{13}$, are mainly due to
the variation in $\delCP$, while the dependence on $E_\mu$ may be
discerned from the variation across the three plots in the panel.
For each baseline, the upper edge of the band gives the lowest value
of $\theta_{13}$ for which hierarchy can be determined at 5$\sigma$ in the most 
conservative case, i.e. irrespective of what the true 
value of $\delCP$ is.
 
The leftmost top panel shows that for $E_\mu$ = 5 GeV, 
the reach of the experiment is optimal at $L \sim$ 2500 km
for most of the $\delCP$ values, however the actual reach depends strongly
on the true $\delCP$. 
At the optimal baseline the reach in $\sin^2\theta_{13}$ varies over 
an order of magnitude, from $\sim 10^{-4}$ to $10^{-3}$.
The maximum reach around this baseline is for $\delCP$(true) = $3\pi/4$, 
which is consistent with the analytic estimate obtained in \cite{dgrprl}.
For increasing parent muon energies, the optimal baseline typically
increases. For longer baselines the maximum of NH and minimum of IH 
shifts to higher energies. Although the condition 
$E_{\rm max}^{\rm NH} = E_{\rm magic}^{\rm IH}$ is not exactly satisfied, the 
broader width of the oscillation curve for higher baselines 
contributes to give an enhanced sensitivity. 
The widths of the bands in the top panel 
are rather conservative, since they include all true values of $\delCP$. 

In reality there is a unique true value of $\delCP$, which controls
the reach for $\sin^2\theta_{13}$ for a given baseline.
Since this true value is unknown in the analysis it has
to be kept free. In order to understand 
what are the specific values of $\delCP$ for which 
hierarchy can be determined at a particular baseline, 
we also present the Type-B plots in the middle panel of 
Fig.~\ref{hierarchy-Lopt}, plots for 
fixed $\theta_{13}$ values in the $\delCP -L$ plane.

The middle-panel plot for 5 GeV muon energy shows that if 
$\sin^2\theta_{13}$ is 0.056, then it is possible to determine 
hierarchy at 5$\sigma$ level for all values of $\delCP$ 
for $L \geq 700$ km. For $\sin^2\theta_{13} =0.01$, one needs 
to go beyond 1200 km if $\delCP > 3\pi/2$. 
For $\sin^2\theta_{13} = 0.001$, there is some sensitivity to hierarchy 
for $L \simeq 1100$ km, but only for $\delCP$ 
close to $\pi/2$. 
Hierarchy sensitivity at 1300 km for this value of 
$\theta_{13}$ exists only for $\sim 32\%$ of the possible $\delCP$ values.
The baseline that offers sensitivity to the largest range of
$\delCP$ values at such low $\theta_{13}$ is 1800--2500 km.
As we go beyond 3000 km the 
range of $\delCP$ values for which hierarchy can be 
determined becomes much smaller for $E_\mu$ = 5 GeV. However
if the energy is increased, then higher baselines can 
determine hierarchy for a wider range of $\delCP$ values. 
One striking feature, seen from these plots, is that there is a small range 
of $\delCP$ values around $3 \pi/2$ for which hierachy 
cannot be determined by LENF for $\sin^2\theta_{13} = 0.001$. 
This can be remedied by a combination with another experiment
that is sensitive to the region around $\delCP = 3\pi/2$.

Another uncertainty in the hierarchy sensitivity 
of a given experimental setup can come from $|\dm{31}|$, 
as the magic energies depend on it. 
To illustrate this dependence, we show {Type-C plots} 
in the bottom panel of Fig.~\ref{hierarchy-Lopt} 
for three fixed $\delCP$(true) values: $\delCP$(true) = 0 (no CPV), 
$\pi/2$ (maximum CPV), and $3\pi/4$ (intermediate CPV). The true 
value of $|\dm{31}|$ is varied over its $3\sigma$ range, giving rise 
to bands with a finite width. Each plot shows the bands for the 
three representative $E_\mu$ values, stated before. 

The widths of the bands in the bottom panel 
of Fig.~\ref{hierarchy-Lopt} show that the 
dependence of hierarchy sensitivity on $|\dm{31}|$ is much smaller 
compared to that on the true value of $\delCP$. 
From the left most figure in the panel,
we see that for no CPV, the best hierarchy sensitivity is obtained 
in a rather broad baseline regime 1500 -- 3500 km, though
for the worst-case values of $|\dm{31}|$, the range around $2500$ km
is preferred, where the error due to $|\dm{31}|$ is also small.
If CPV is maximum ($\delCP = \pi/2$, central panel) or has the chosen 
value of $3\pi/4$ (right panel), the maximum hierarchy sensitivity is 
for a comparatively narrower, but still wide, baseline range $\sim$ 
2300 -- 3500 km, near the bimagic baseline.
At the worst-case values of $|\dm{31}|$, a baseline $\sim 2500$ km
is preferred for $E_\mu$ = 5 GeV, while it shifts to $\sim$3000 km for 
higher $E_\mu$ values. 
Among the three $\delCP$ values considered, the sensitivity to hierarchy 
is observed to be the worst for the scenario with no CPV, however as
can be seen from the top panel, different values
of $\delCP$ are the worse-case scenarios for different baselines.
As the parent muon energy increases, the optimal baseline shifts to 
higher values as expected.
The sensitivity is seen to increase when $E_\mu$ increases from 5 to
7.5 GeV, but thereafter it seems to saturate. This feature will 
be discussed in detail in the next sections.

Our results in this section clearly indicate that if hierarchy
sensitivity is considered as the performance indicator, 
and if $\sin^2\theta_{13}~\ls 0.001$ and $E_\mu = 5$ GeV, 
then the hierarchy can be determined for a 
larger ($\sim 86\%$) fraction of possible $\delCP$ values at 2540 km,
the bimagic baseline. This fraction increases for higher $E_\mu$
values. For $\sin^2 \theta_{13} > 0.01$, hierarchy determination is
possible for all values of $\delCP$, as long as $E_\mu > 5$ GeV and 
$L > 1300$~km.

\subsection{$\theta_{13}$ discovery} 
\label{sec:th13-Lopt}

\begin{figure*}[t]
\begin{center}
\includegraphics[angle=0,width=1.8\columnwidth]{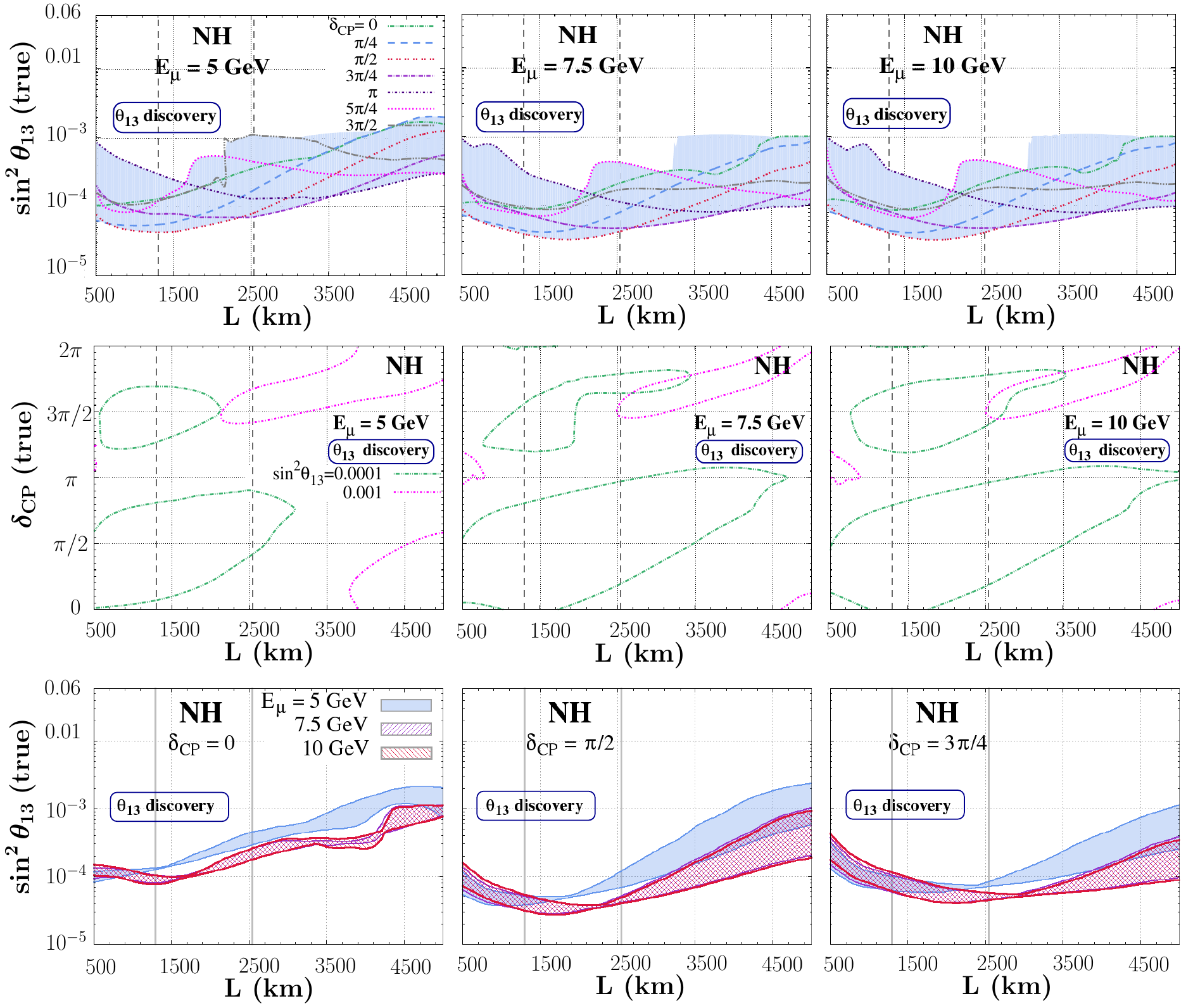} 
\caption{(Color online) 5$\sigma$ reach in $\theta_{13}$ discovery 
for fixed muon energies, as a function of baseline, assuming the true 
hierarchy to be NH. 
The same conventions as in Fig.~\ref{hierarchy-Lopt} are used. 
In the middle panel, for
$\sin^2\theta_{13} = 0.001$, the discovery of $\theta_{13}$ is
possible for regions outside the dashed (magenta) contour. For
$\sin^2\theta_{13} = 0.0001$, the discovery of $\theta_{13}$ is
possible for the  area inside the dot-dot-dashed (green) contours.
\label{th13-Lopt}}
\end{center}
\end{figure*}

In this section, we apply the same analysis techniques as the last
section to the discovery potential of $\theta_{13}$. Note that though
a $\sim 2.5\sigma$ evidence for nonzero $\theta_{13}$ has recently been 
claimed by experiments \cite{t2k-th13,minos-th13} and the global fit to the
neutrino mixing parameters \cite{latest-fit-lisi,latest-fit-schwetz}, 
the jury is still out on this 
and it is quite possible that the value of $\theta_{13}$ is much smaller.
Also, since we are looking towards a long-term experiment, we
should not be satisfied with a $3\sigma$ evidence, but should try to
gauge the potential of an experiment for a definitive, $5\sigma$ 
discovery\footnote{ 
The implications of the recent Daya Bay \cite{dayabay-th13} and
RENO \cite{reno-th13} results are discussed towards the end of the paper.}. 
Fig.~\ref{th13-Lopt} shows the 5$\sigma$ reach for $\theta_{13}$ 
discovery for three different parent muon energies, in terms of 
the three types of plots, {Type-A, B and C, as} mentioned above.

The widths of the bands in the top panel show that $\theta_{13}$ discovery 
potential of a baseline depends strongly on the true value of $\delCP$.
Indeed for any baseline, the best-case and the worst-case values of
$\delCP$ make a difference of almost an order of magnitude in the
corresponding reach for $\sin^2\theta_{13}$.
Shorter baselines are seen to be generally preferred for $\theta_{13}$
discovery, as compared to those preferred for hierarchy determination.
For example, for $E_\mu$ = 5 GeV, the optimal baseline is $\sim 800-1600$ km,
while for higher parent muon energies, it moves to $\sim 1500-2500$ km.
At all these energies, it is observed that the sensitivity is maximum
when $\delCP \approx \pi/2$. 
With the worst-case values of $\delCP$, the optimal baseline stays
near 1500 -- 2000 km in the whole energy range.

In the middle panel of Fig.~\ref{th13-Lopt} we present the {Type-B} 
plots. For $\sin^2\theta_{13} = 0.001$, the discovery of $\theta_{13}$ 
is possible for regions outside the magenta (dashed) contour. 
Thus for $L < 2100$ km, $\theta_{13}$ can be discovered irrespective of 
$\delCP$ for this value of $\sin^2\theta_{13}$. 
Higher $\theta_{13}$ values can be discovered in the whole 
$\delCP - L $ plane. 
For $\sin^2\theta_{13} = 0.0001$, the discovery of $\theta_{13}$
is possible for the area inside the dot-dot-dashed green contours i.e. 
only for shorter baselines and a limited range of $\delCP$. 
For higher baselines like $L \sim 3000$ km, $\theta_{13}$ discovery is 
possible only if $E_\mu$ is high and $\delCP$ has values close to 
$\pi/2$ or $7\pi/4$.
In general for such small values of $\sin^2\theta_{13}$, the
baseline of $\sim$1500 km seems to have $\theta_{13}$ sensitivity for 
a wider range of $\delCP$ values.

In the bottom panel of Fig.~\ref{th13-Lopt}, we 
present the { Type-C} plots for $\theta_{13}$ sensitivity. 
It illustrates the dependence of the optimal baseline on
the true value of $\delCP$: while for $\delCP=0$ the optimal baseline
seems to be $\sim 1300-1500$ km, for $\delCP=\pi/2$ it is $\sim 1500-2000$
km and for $\delCP=3\pi/4$, the most efficient baseline is $\sim 2500$ km.

\begin{figure*}[]
\begin{center}
\includegraphics[angle=0,width=1.8\columnwidth]{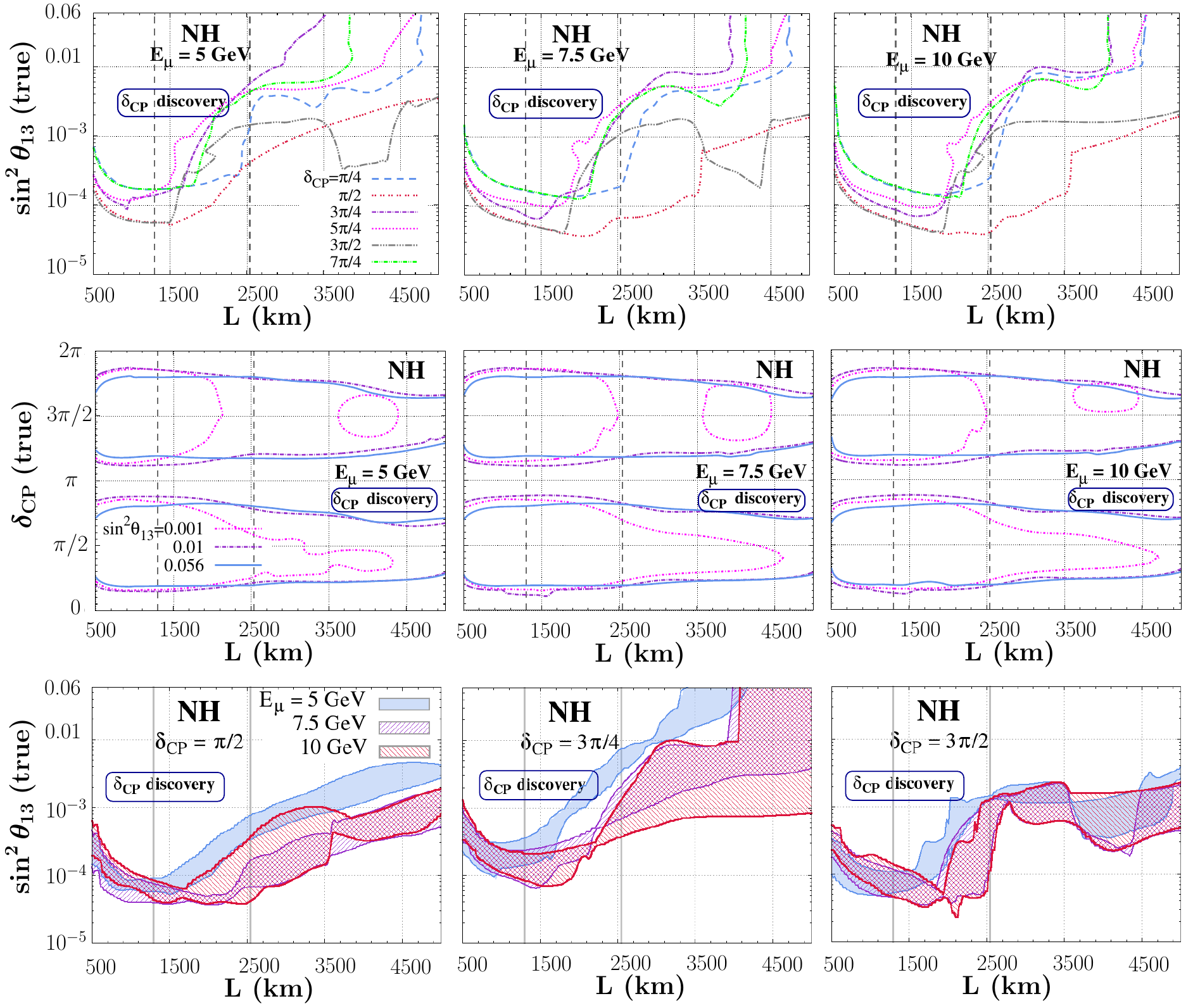} 
\caption{(Color online) 5$\sigma$ reach in $\delCP$ discovery 
for fixed muon energies, as a function of baseline, assuming the true 
hierarchy to be NH. 
The same conventions as in Fig.~\ref{hierarchy-Lopt} are used. In the middle panel,
$\delCP$ discovery is possible for enclosed regions that do not
include $\delCP$ =0 or $\pi$.}
\label{cpv-Lopt}
\end{center}
\end{figure*}

To summarize, if $\theta_{13}$ discovery is chosen as the 
performance indicator, baselines in the region 1500 -- 2000 km 
seems to work best, the actual value of the optimal baseline being 
dependent on the true value of $\delCP$. 
However if $\sin^2\theta_{13} > 0.001$, then a LENF with
$E_{\mu}$ in the range 5 - 10 GeV can discover 
$\theta_{13}$ at 5$\sigma$ for any baseline in the range 500--2500 km 
for almost all values of true $\delCP$.

\subsection{$\delCP$ discovery}
\label{sec:delcp-Lopt}

Next we move to the $\delCP$ discovery potential in this section.
The top panel of 
Fig.~\ref{cpv-Lopt} shows the 5$\sigma$ reach of $\sin^2\theta_{13}$ 
for $\delCP$ discovery, given a true $\delCP$ value.
True values of all other parameters have been fixed to values 
mentioned in Eq.~(\ref{simulation-para}), while the true value of $\delCP$ 
is chosen as $\pi/4$, $\pi/2$, $3\pi/4$, $5\pi/4$, $3\pi/2$ and $7\pi/4$. 
We do not choose 
$\delCP$(true) too close to zero or $\pi$ since the discovery potential is 
expected to be low. Finally $\chimin$ is obtained by marginalizing over 
all parameters except $\delCP$, for each $\sinsq{13}$(true). 
Errors on different parameters are taken to be the same as stated in 
Section~\ref{sec:hierarchy-Lopt}.

It is observed that the best-case sensitivity for $\delCP$ is when
$\delCP$ is near $\pi/2$. For smaller baselines $\delCP = 3 \pi/2$
also has similar sensitivity, while between 3500 -- 4500 km and for
energies 5 and 7.5 GeV the sensitivity is better.
It is expected that $\delCP = \pi/2$ and $3\pi/2$ should produce 
the best-case sensitivity for $\delCP$ since it is at
these values that the CP violation is maximum. Typically for all
$\delCP$ values and all parent muon energies, the baselines 
$\sim 800-2000$ km seem to be the most efficient. 
For $\delCP=3 \pi/4, 5\pi/4, 7 \pi/4$, on the other hand, the sensitivity 
is much worse. Nevertheless, baselines of $\lesssim 2000$ km are
preferred.

The middle panel of Fig.~\ref{cpv-Lopt} gives the sensitivity to $\delCP$ 
as a function of baseline for three fixed values of $\sin^2\theta_{13}$ 
which are 0.056, 0.01 and 0.001 and three values of energy 
$E_\mu = 5, 7.5$ and 10 GeV. 
These are representative values and the results for 
intermediate values of $\sin^2\theta_{13}$ can be adjudged.
The figure shows that for $\sin^2 \theta_{13} > 0.01$, 
if $\delCP$ lies in the ranges $(0.3 - 0.7)\pi$ or $(1.3 - 1.7)\pi$, 
it is possible to discover nonzero $\delCP$ at 5$\sigma$ 
for all baselines and all energies considered here, though
smaller baselines are a bit more efficient.
For lower $\theta_{13}$ values, only shorter baselines, $L \lesssim 2000$ km, 
have the possibility of detection of CP violation for most of the
$\delCP$ range.
Longer baselines need higher muon energies and specific $\delCP$
values near $\pi/2$, $3\pi/2$ in order to achieve the task.

The bottom panel of Fig.~\ref{cpv-Lopt}
gives the { Type-C} plots for $\delCP$ discovery for 
three fixed values -- $\pi/2$, 3$\pi/4$, 3$\pi/2$. 
This figure corroborates that the sensitivity to CP violation discovery
is highest for $\delCP = 3 \pi/2$. 
The distance at which the best sensitivity is 
reached is $\sim$1700 km (for $\delCP = \pi/2$) 
and $\sim$2100 km ( for $\delCP = 3 \pi/2$). 
For $\delCP = 3 \pi/4$ the range 1500 - 2500 km have best sensitivity. 
In general the sensitivity is better at higher energies.

To summarize, the $\delCP$ discovery potential at various baselines 
depends on the true $\sin^2\theta_{13}$. 
For $\sin^2\theta_{13}$ upto 0.001, $\delCP$ that is not too close to 
0 or $2 \pi$ can be discovered at 5$\sigma$ by the LENF setup for 
baselines 500 -- 2000 km. 
In general lower baselines can access a wider range of $\delCP$ values.

\section{Optimizing parent muon energy $E_\mu$} 
\label{sec:Eopt}

As seen in the last section, the optimal values of the baselines
depend on the parent muon energies in addition to the true values
of neutrino mixing parameters.
In this section, we choose three representative values for the baseline
-- 1500, 2500 and 3500 km -- and perform optimization over the
parent muon energies.
We use three different performance indicators as before: 
hierarchy determination, $\theta_{13}$ discovery and $\delCP$ discovery. 
The parent muon energy range considered here is 2 -- 16 GeV, covering the 
range of the proposed LENFs.
We present only plots of {Type-A}, since most of the features of the energy 
dependence may be obtained through interpolation using the
representative values $E_\mu = $5, 7.5 and 10 GeV chosen in the last section.

\begin{figure*}[t!]
\begin{center}
\includegraphics[angle=0,width=1.8\columnwidth]{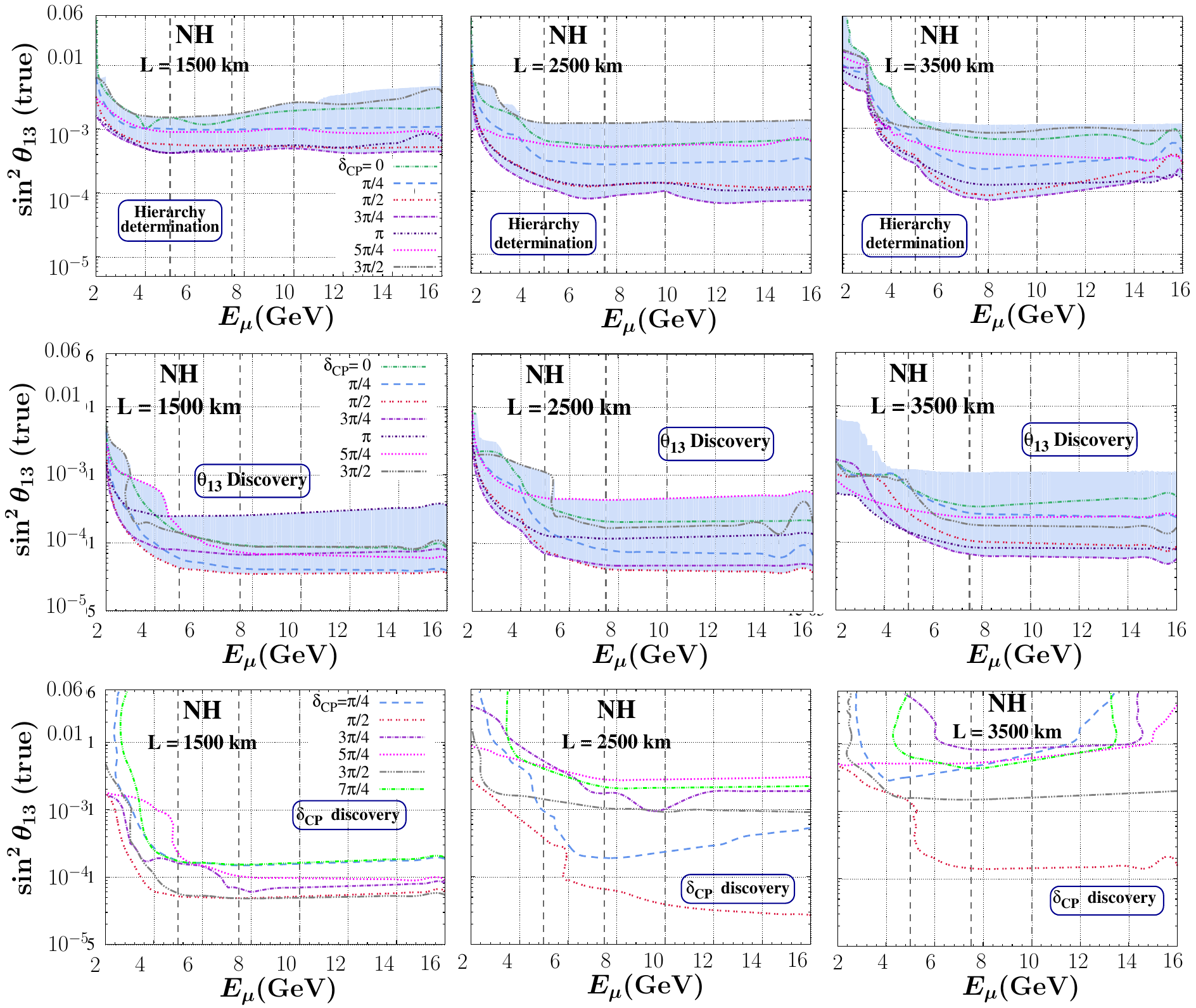} 
\caption{(Color online) 
5$\sigma$ reach in $\sin^2\theta_{13}$ 
for (top panel) hierarchy determination, (middle panel) $\theta_{13}$
discovery, and (bottom panel) nonzero $\delCP$ discovery,
as a function of $E_\mu$, for fixed $\delCP$ values. 
NH is taken as the true hierarchy. 
All parameters except $\delCP$ are fixed at values quoted 
in Eq.~(\ref{simulation-para}). 
The exposure is taken to be 2.5 years of running with each polarity. 
The dark vertical lines correspond to energies 5 GeV, 7.5 GeV and 10 GeV.
}
\label{fig-Eopt}
\end{center}
\end{figure*}

\subsection{Hierarchy determination}
\label{sec:hierarchy-Eopt}

The top panel of Fig.~\ref{fig-Eopt} shows the 5$\sigma$ hierarchy 
determination range for the three chosen benchmark baselines.
For each chosen true value of $\sinsq{13}$ and $E_\mu$, we obtain $\chimin$ 
by marginalizing over all parameters assuming IH, while the true hierarchy 
is taken to be NH. All the parameters, except $\delCP$, are fixed to 
the values quoted in Eq.~(\ref{simulation-para}) and the 
band is generated by varying $\delCP$ 
in the full range $[0, 2\pi]$. Errors on different parameters have been 
taken to be the same as described in Section~\ref{sec:hierarchy-Lopt}.

Similar to the case of baseline optimization in Fig.~\ref{hierarchy-Lopt}, 
the large width of the band indicates the strong dependence of hierarchy 
determination potential for a given experimental setup on the true value 
of $\delCP$. 
As far as the dependence on $E_\mu$ is concerned, 
the sensitivity increases with $E_\mu$ at low $E_\mu$ values, and saturates 
at a certain value of $E_\mu$ beyond which there is virtually no change
in the sensitivity. Let us refer to this energy as the 
saturation energy, $\Esat$. 
It is observed that $\Esat$ increases with increasing baseline, 
the values being $\Esat \sim 5$ GeV for 1500 km, $\Esat \sim 7$ GeV 
for 2500 km and $\Esat \sim 10$ GeV for 3500 km. 
The sensitivity reached at $\Esat$ also increases with the baseline.

The saturation behavior indicates that one will tend to get
significantly better sensitivity with increasing $E_\mu$ and baseline 
till some limit, beyond which the gain may not be worth the
increased acceleration required.
We observe that $\Esat$ depends significantly on $\delCP$ as well,
however this dependence is not a straightforward one.
The sensitivity at a baseline of 2500 or 3500 km is always 
better than that at 1500 km, sometimes by up to an order of magnitude, 
for energies beyond $\sim$ 6 GeV. 
The performances at 2500 and 3500 km are comparable: the latter is
marginally better, but requires higher $E_\mu$ to achieve its full potential.

Thus if hierarchy determination is considered as the performance indicator, 
then a LENF can determine hierarchy for 
$\sin^2\theta_{13}$ as low as $0.002$ at a baseline 
$L \sim 2500$ km for $E_{\mu} > 5 GeV$, for any $\delCP$(true). 
If true $\sin^2\theta_{13}$ happens to be large, then 
shorter baselines and smaller energies would be sufficient.

\subsection{$\theta_{13}$ discovery}
\label{sec:th13-Eopt}

In this section we present the result of our study of $\theta_{13}$ 
discovery potential as a function of the parent muon energy $E_\mu$, 
for the three benchmark baselines.
The middle panel of Fig.~\ref{fig-Eopt} shows the 5$\sigma$ contours when 
$\delCP$(true) is varied in $[ 0, 2\pi ]$ and also for seven chosen true 
values of $\delCP$.
True values of all other parameters other than the displayed 
ones have been fixed to the values mentioned in 
Eq.~(\ref{simulation-para}), 
and $\chimin$ is obtained by marginalizing over all except $\theta_{13}$, 
with the errors stated in Section~\ref{sec:hierarchy-Lopt}.

These figures also confirm that the discovery potential of any experimental 
setup depends crucially on the knowledge of $\delCP$.
For 1500 km, the sensitivity is the best when $\delCP \approx \pi/2$, 
while for higher baselines $\delCP \approx 3\pi/4$ has comparable or slightly 
better sensitivity, however the conservative reach for $\theta_{13}$ may be up to
an order of magnitude worse.
Larger baselines require larger muon energies to get an
equivalent sensitivity, however with such a larger $E_\mu$,
their reach in $\theta_{13}$ may be better.

The saturation behavior as in the case of hierarchy determination -- 
i.e. the sensitivity increases with $E_\mu$ till 
a saturation value $\Esat$ -- is observed even for $\theta_{13}$ discovery 
for most $\delCP$ values.

\subsection{$\delCP$ discovery}
\label{sec:delcp-Eopt}

The analysis of the optimal parent muon energy for $\delCP$ discovery
is carried out on the same lines as that in Sec.~\ref{sec:delcp-Lopt}.
The plots in the bottom panel of Fig.~\ref{fig-Eopt} show the 5$\sigma$ 
discovery potential contours as a function of $E_\mu$, for six chosen 
true values of $\delCP$, for each of the three representative baselines.

It is observed that in general the sensitivity to $\delCP$ discovery 
decreases as the baseline increases. For most values of $\delCP$ the baseline 
of 1500 km with $E_\mu$ = 6 -- 7.5 GeV is seen to give the best sensitivity 
in Fig.~\ref{fig-Eopt}, while for $\delCP = \pi/2$ a baseline of 2500 km 
and energy $\sim 12$ GeV seem to do better. For all the three chosen 
baselines better sensitivity to $\delCP$ comes beyond $E_\mu = 5$ GeV.


\section{Optimal parent muon energy for a given baseline}
\label{emu-given-baseline}

The two design parameters of an LENF that can in principle be controlled
are the baseline and the parent muon energy that determines the 
neutrino fluxes at the source. Ideally one can look for the optimal 
combination of both of these, however it may not always be practical.
In particular, considerations behind choosing a baseline involve
factors like the location suitable for an accelerator and a location
where a undergraound laboratory for neutrino detection can be built.
Apart from scientific considerations, this involves geography, economics 
as well as sociology.
A survey of such pairs of locations was recently carried out 
\cite{agarwalla-optimal}. 
Given a baseline, the choice of the parent muon energy is relatively 
straightforward, limited mainly by technological considerations. 
The energy of the muon beam can then be chosen based on the optimality 
analysis.

As has been noticed multiple times in the previous two sections,
the performance at a baseline typically increases with increasing
parent muon energy $E_\mu$ (in the LENF range), till it saturates at
a particular value $\Esat$. Given that increasing the muon energy
is associated with additional costs, a desirable thing to do
would be to choose the value of $E_\mu$ at or near the value of
$\Esat$. In this section, we obtain the $\Esat$ values for
arbitrary baselines using a simple approximation. It turns 
out that this approximation matches the numerical $\Esat$ values
observed in the previous section.

We expect that the main factor influencing the efficiency at a given energy 
will be the number of wrong-sign muon events. While the actual numbers
will depend on the detector characteristics, we estimate this
efficiency simply through the ``quality factor''
\beq
Q \equiv \frac{\int \Phi_{\nu_e} P_{e \mu} \sigma_{\nu_\mu} dE_{\nu_e}}{\int 
\Phi_{\nu_e} \sigma_{\nu_\mu} dE_{\nu_e}}
\label{Q-def}
\eeq
where $\Phi_{\nu_e}$ is the flux of $\nu_e$ at the source and $\sigma_{\nu_\mu}$
is the total charged-current cross section of $\nu_\mu$ at the detector.
We have normalized the events to a complete $\nu_e \to \nu_\mu$
conversion. 
Since the parent muon energy at the source determines the neutrino
spectra at the source completely, and the baseline determines the
oscillation probability completely (modulo our knowledge about the
mixing parameters), the quality factor is known for a energy-baseline
combination. The optimal energy $\Eopt$ is the one for which this
quality factor is maximum. There will be a spread in $\Eopt$ due to 
the uncertainties in mixing parameters, mainly $\theta_{13}$, 
$|\dm{31}|$ and $\delCP$.

\begin{figure}[t]
\includegraphics[height=4.6cm, width=6.8cm]{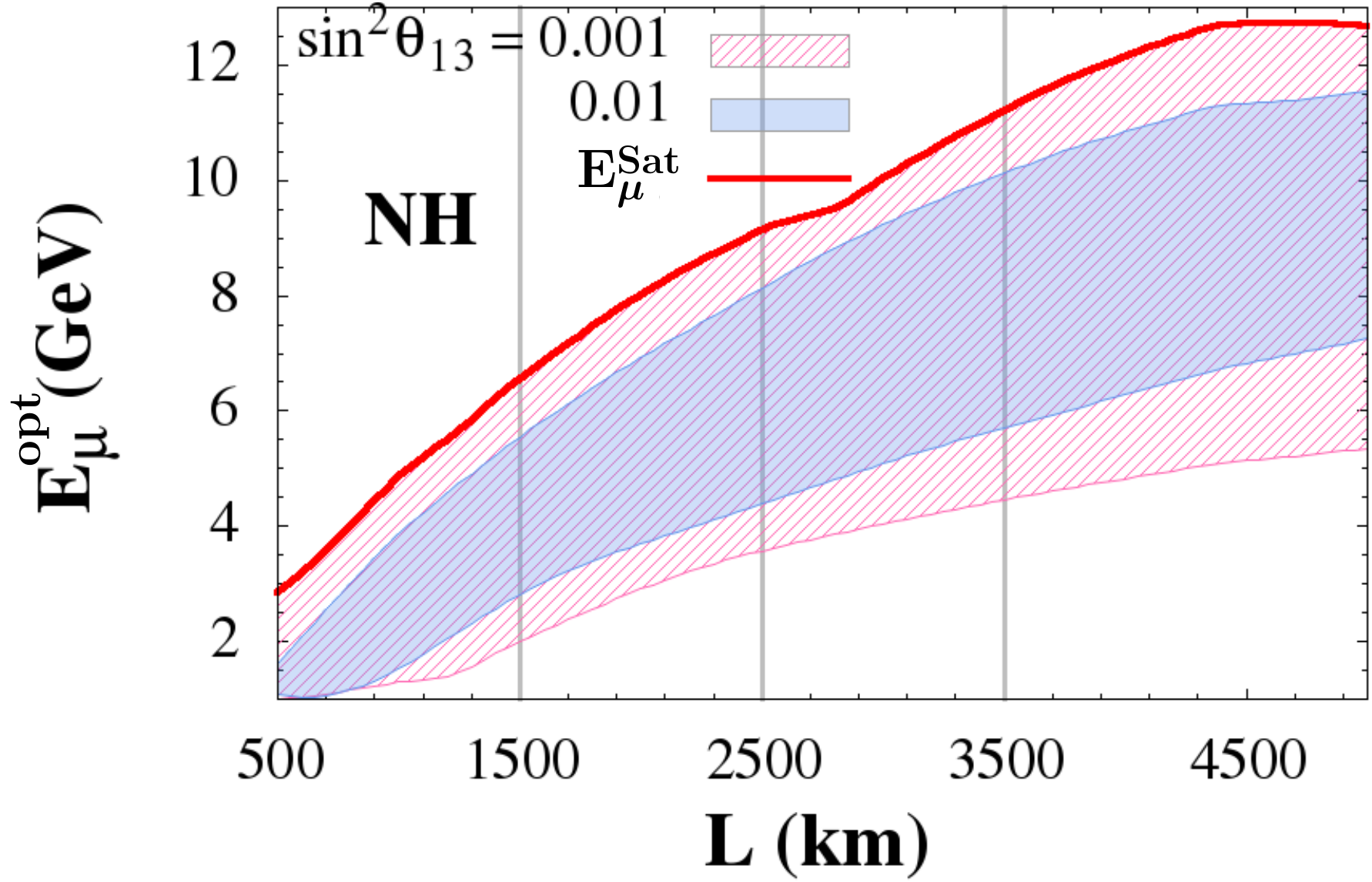}
\caption{(Color online) 
The optimal value $\Eopt$ as a function of the baseline, for
normal hierarchy and with $\sin^2\theta_{13} = 0.001$ and $0.01$. 
}
\label{fig:L-E-optimize}
\end{figure}

Fig.~\ref{fig:L-E-optimize} shows the $\Eopt$ values as a function of 
baseline $L$, for $\sin^2\theta_{13}$ = 0.001, 0.01. The 
width is obtained due to variation of $\Eopt$ as $|\Delta m^2_{31}|$ 
is varied in the current 3$\sigma$ range and $\delCP$ in [0, 2$\pi$].
This gives a good estimation of what $E_\mu$ values will be optimal
corresponding to a given baseline,
and can work as a rough guideline for determining the optimal 
range of $E_\mu$, once the baseline $L$ has been determined from
other considerations. Then the upper limit can be 
used to estimate $\Esat$ for a given baseline. Going beyond 
$\Esat$ would not improve the performance of the setup.

\begin{figure*}[]
\begin{center}
\includegraphics[angle=270,width=1.8\columnwidth]{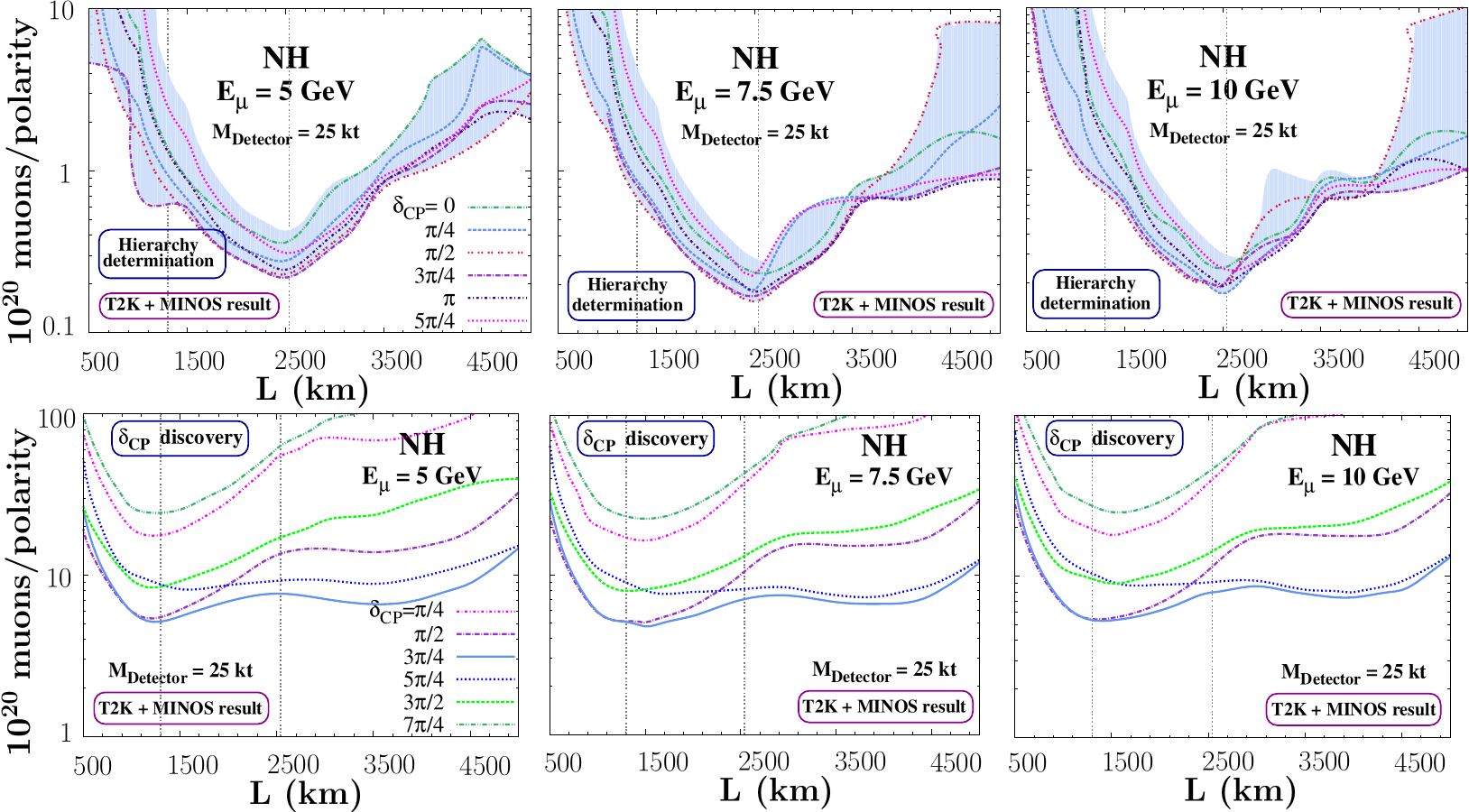}
\caption{(Color online) Exposure required for 5$\sigma$ hierarchy determination 
(top panel) and $\delCP$ discovery (bottom panel), if $\theta_{13}$ is large. We have taken 
$\sin^2\theta_{13} \in$ [0.013, 0.028] \cite{latest-fit-lisi}. NH is taken 
as the true hierarchy, and all parameters except $\delCP$ has been 
fixed at the values stated in Eq.~(\ref{simulation-para}).
\label{fig:contour-larget13}}
\end{center}
\end{figure*}

\section{If $\theta_{13}$ is large}
\label{large-t13}

The indications of a large $\theta_{13}$ value from the experiments
\cite{t2k-th13,minos-th13} and the global fit to neutrino data
\cite{latest-fit-lisi,latest-fit-schwetz} is a good news for 
the measurements of other
neutrino parameters too. In particular, it makes
it easier to determine the mass hierarchy, and makes the 
measurement of $\delCP$ possible\footnote{ Also, see the Note added
at the end of this paper for comments on the recent Daya Bay 
\cite{dayabay-th13} and RENO \cite{reno-th13} results.}.

The figures presented in the previous section already give us some 
idea about what can be the optimal energies and baseline for hierarchy 
and $\delCP$ for $\sin^2\theta_{13}$ in the above range. 

If $\sin^2\theta_{13}$ is close to its present best-fit, then 
the LENF set up considered in this paper can determine hierarchy 
irrespective of what is the true $\delCP$ for baselines $\gs$ 1000 km,
with a 2.5 years exposure with each of the polarities and 
$5 \times 10^{21}$ useful muon decays, as shown in 
Figure~\ref{hierarchy-Lopt}.
If true $\delCP$ is not too close to $3 \pi/2$ then even 
baselines $\gs$ 700 km would be sufficient for this purpose. 
The task of $\delCP$ determination will also be easier:
$\delCP$ in the range $ (0.2 - 0.8) \pi$ and $ (1.2 - 1.8) \pi$ can 
be determined at 5$\sigma$ for baselines of 500--2000 km for 
$E_{\mu}$ in the range 5-7 GeV, as can be seen from 
Figure~\ref{cpv-Lopt}.

We can then be more ambitious and try to optimize the setup in
order to get a $5\sigma$ determination of hierarchy and $\delCP$
with the minimum exposure. 
This is shown in Fig.~\ref{fig:contour-larget13}, where 
we take the range of $\sin^2\theta_{13}$ to be [0.013, 0.028], the
$1\sigma$ range given in \cite{latest-fit-lisi}, in the prior and 
then marginalized over $\theta_{13}$.
 
It is clear from the top panel of the 
Fig.~\ref{fig:contour-larget13} that for any parent muon energy, the baseline 
$L \sim 2500$ km is the optimal one for the determination of mass hierarchy. 
As mentioned before, this is near the bimagic baseline indicated in 
\cite{dgrprl}.
For a baseline of 1500 or 3500 km, the exposure needed may be
an order of magnitude larger, depending on the $\delCP$ value.

The bottom panel of Fig.~\ref{fig:contour-larget13} shows that 
for any $E_\mu$ a baseline $\sim$1300 km will show the best 
sensitivity for $\delCP$ discovery, for most of the true values of 
$\delCP$. However, the ``bimagic baseline'' 2540 km will have 
comparable sensitivity with an exposure $\sim$1.5 times larger. If 
$\delCP$(true) happens to be close to $5\pi/4$, $\sim$2500 km may 
show a better sesitivity compared to 1300 km. 
Note that the exposure needed for $\delCP$ discovery at 5$\sigma$ 
is typically one order of magnitude higher than that for 
hierarchy determination.

\section{Summary and concluding remarks} 
\label{concl}

The neutrino mixing angle $\theta_{13}$, the CP-violating phase 
$\delCP$ and the hierarchy of neutrino mass eigenstates are the
three quantities whose measurements are crucial in order to complete
and confirm our current picture of neutrino mixing and oscillations.
The next generation of long-baseline neutrino experiments therefore 
justifiably consider these measurements as their primary goals. 

In this article, we go beyond the conventional
beam experiments and try to optimize the setup for a 
low-energy neutrino factory (LENF)
-- where the energy of the parent muon is less than 
$\sim 15$ GeV, and consider a magnetized totally-active 
scintillator detector (TASD)
that can identify muon charge, for these measurements.
In the worst case scenario i.e., $\theta_{13}$ turning out to be smaller than
the reach of the current reactor and superbeam experiments, such setups can be
used as discovery machines. On the other hand if $\theta_{13}$ is confirmed to
be in the currently indicated range then one can use such machines for
precision studies with the aim to measure the three quantitites at the
5$\sigma$ level. In this paper we have focussed on the discovery 
potential.   

We note that some of the parameters that determine the efficiency of a neutrino
factory are beyond our control: the probability $P_{e\mu}$ that controls
this efficiency depends crucially on $\theta_{13}$ and $\delCP$. 
However there are two parameters in our control: the baseline and 
the energy of the parent muon. It is with respect to these
parameters that we perform our optimization.

It is of course obvious that the value of $\theta_{13}$ would determine
how efficient an experiment is. And that is true more or less in 
a straightforward way. In the absence of matter effects, 
$P_{e\mu}$ is proportional to $\theta_{13}^2$. 
In the presence of matter effects, $\theta_{13}$ affects the 
probability linearly, thus influencing the identification of mass hierarchy.
The term in the probability that involves $\delCP$ is also
linear in $\theta_{13}$. 
Thus at shorter baselines (low matter effects), the value of $\theta_{13}$
affects the measurement of $\theta_{13}$ itself through a quadratic term,
while at larger baselines, it affects the probability through a
term that contains the product of itself with the matter effects.

In addition to $\theta_{13}$, we find that even the true value of
$\delCP$ can have a large impact on the reach of an LENF
experiment. Indeed, the reach can change by up to an order of magnitude 
or more depending on the true value of $\delCP$.
The actual value of $|\dm{31}|$ also affects the efficiency,
albeit to a smaller extent. 
Given that the actual value of $\delCP$ is crucial, we present our
results in a way that bring out the impact of this parameter. We 
present three types of plots:
(i) Type-A: plots in the $\sin^2\theta_{13}-L$ plane for fixed values of
$|\Delta m^2_{31}|$ and $\delCP$ varying in the range [0, $2\pi$], 
(ii) Type-B: plots in the $\delCP- L $ plane for fixed values of 
$\sin^2\theta_{13}$ and $|\Delta m^2_{31}|$,
(iii) Type-C: plots in the $\sin^2\theta_{13}-L$ plane for fixed values of 
$\delCP$ and varying $|\Delta m^2_{31}|$ in its current 3$\sigma$ 
allowed range.

Most of the earlier work with the optimization of LENF presented
the results in terms of ``fraction of $\delCP$'' where the experiment
was successful. However, it is crucial to know the exact range of 
$\delCP$ where a certain measurement is possible, especially when one
is thinking about combining results from two complementary experiments:
these experiments should be sensitive to complementary ranges of
$\delCP$.

Our detailed observations may be found in the main body of this text.
We would like to attract the reader's attention to some salient
features. The following observations refer to 2.5 years of running
with each muon polarity, with $5 \times 10^{21}$ useful muon decays
per year.\\
(i) For $\sin^2 \theta_{13} \gtrsim 10^{-2}$, mass hierarchy can be
determined to $5\sigma$ at almost all the baselines $> 1000$ km.
For lower $\theta_{13}$, baselines $\sim 2500$ km achieve the 
task for the largest fraction of $\delCP$, however even they may 
fall short when $\delCP$ is near $3\pi/2$.\\
(ii) Any baseline $\lesssim 2000$ km can discover 
$\theta_{13}$ if $\sin^2\theta_{13} \gtrsim 10^{-3}$. 
For smaller $\theta_{13}$, the choice 
of optimal baseline depend on the actual value of $\delCP$.
However for majority of values of $\delCP$ the baseline range
1300 - 2500 km can be termed as optimal. 
\\
(iii) The discovery of $\delCP$ is naturally harder when its value
is near $0$ or $\pi$. However if $\sin^2\theta_{13} \gtrsim 0.01$,
the discovery is possible for a wide range of $\delCP$ values
centered at $\pi/2$ and $3\pi/2$. The range typically decreases
with the increase of baseline.  \\
(iv) At a given baseline, when the parent muon energy is increased,
the performance typically increases up to some energy and then
remains the same. We term this as the saturation behavior.

As can be gathered from above, there does not exist a unique ``optimal''
baseline or muon energy for all performance indicators.
The baseline determination depends, in addition to scientific merit, 
also on geography, economics and sociology. Once that is determined, 
the optimization of muon energy involves mainly scientific and technical 
considerations. The saturation behavior mentioned above indicates that, 
given a baseline, it is preferable to have the parent muon energy as close
to the saturation energy as possible, in order to avoid increasing the
parent muon energy unnecessarily.
While the saturation energy may be determined through a detailed
simulation for a given baseline (as we have done for three benchmark
baselines in this article), we have tried to come up with a simple
criterion that can motivate this behavior and determine the saturation
energy. We conjecture that the most efficient energy for a given baseline
simply depends on the number of wrong-sign muons at the detector.
This gives us a range of optimal muon energies for a given baseline,
the range depending on the actual values of mixing parameters. 
This conjecture is vindicated a posteriori by the saturation energies
obtained at the benchmark baselines.
This simple-minded analysis gives an intuitive understanding of 
the relationship between baseline and the optimal energy.

With the recent indications of a large value of $\theta_{13}$, 
the $5\sigma$ measurement of this quantity may already be within
our grasp with the current experiments.
Then the neutrino factory experiments have to aim higher, and
the question to ask is what kind of setup will give us the $5\sigma$
determination of mass hierarchy and CP violation with the least
amount of exposure needed. 
Our analysis indicates that for the mass hierarchy, the baseline of
$\sim 2500$ km would perform the best at any parent muon energy.
On the other hand, the measurement of CP violation is the most
efficient around a baseline of $\sim 1500$ km at all parent muon
energies, for most of the $\delCP$ values.

A few comments are in order. We have presented all our results 
assuming the actual hierarchy to be NH.
However note that the probabilities obtained with NH and $\mu+$ beam 
are identical with the probabilities with IH and $\mu^-$ beam, 
and we have taken beams of both polarities with equal exposure.
In LENF, both the polarities of muons can be accelerated 
at the same time, giving alternate bunches of $\mu^+$ and $\mu^-$.
The cross sections of $\nu_\mu$ and $\bar\nu_\mu$ on the nucleons
are also virtually identical at the relevant energies, and hence
our results are valid even for IH.

Another comment is about the measurement of $\delCP$ itself. 
In this article, we have only
considered the discovery potential for CP violation, i.e. we are
interested in finding a nonzero $\delCP$. This task is, naturally,
hard for small $\delCP$ and impossible when $\delCP$=0 or $\pi$.
However with large $\theta_{13}$, a measurement of $\delCP$ may be
possible even if its value is close to the CP-conserving limit.
If indeed the value of $\theta_{13}$ is confirmed to be large,
one can aim to answer more detailed questions like 
the value of $\delCP$, the octant of $\theta_{23}$, 
or the deviation of $\theta_{23}$ from maximality.
These quantities will not be analyzed in this work.

Finally, the most important quantity is the one we have no
control over: the value of $\theta_{13}$. As far as neutrino
factories are concerned, this quantity will determine if
they are going to be discovery machines or precision machines.

\section*{Note added}

While this paper was under review, new results were announced by the
Daya Bay \cite{dayabay-th13} and RENO \cite{reno-th13} experiments
which claimed the discovery of a nonzero $\theta_{13}$ to more than $5\sigma$. 
Indeed the Daya Bay measurement  gives 
$\sin^2\theta_{13}= 0.023 \pm 0.004 \pm 0.001$, while
the RENO experiment gives
$\sin^2\theta_{13}= 0.026 \pm 0.004 \pm 0.004$.
The analysis in this paper has been done assuming that the value of 
$\theta_{13}$ is still unknown, however it stays valid even with a
measured nonzero value. (Of course, the results about $\theta_{13}$
discovery would become redundant.)
Indeed, the projections for the reach of the LENF would become even 
more optimistic, and the results in Sec.~\ref{large-t13} gain even more
significance. 
The range of $\sin^2 \theta_{13}$ taken in the analysis in 
Sec.~\ref{large-t13} is $(0.013, 0.028)$ which overlaps with
the $1\sigma$ range of the Daya Bay and RENO measurements. 
The results in Fig.~\ref{fig:contour-larget13} then indicate that
the baseline of $\sim 2500$ km can yield the mass hierarchy with the
minimum amount of exposure, while the baseline $\sim 1500$ km would
be the optimal for the detection of nonzero $\delta_{CP}$.
Note that with the value of $\theta_{13}$ as large as that measured
by these experiment, the measurement of CP violation at the
LENF would be a real possibility. With a detector capable of 
distinguishing $\mu^+$ from $\mu^-$ (and hence, $\nu_\mu$ from 
$\bar\nu_\mu$), a LENF would then become the front runner in the
race for the first observation of CP violation in the lepton sector.


\section*{Acknowledgments}
We thank S.~ Prakash, S.~ Raut and S.~UmaSankar for useful discussions.
S.R. would like to thank Yuval Grossman for support and hospitality.


\end{document}